\documentclass[lettersize,journal]{IEEEtran}
\pdfoutput = 1
\usepackage{amsmath,amsfonts}
\usepackage{algorithmic}
\usepackage{algorithm}
\usepackage{array}
\usepackage[caption=false,font=normalsize,labelfont=sf,textfont=sf]{subfig}
\usepackage{textcomp}
\usepackage{stfloats}
\usepackage{url}
\usepackage{verbatim}
\usepackage{graphicx}
\usepackage{cite}
\usepackage{amssymb}

\begin{document}

\title{Secure MIMO Communication Relying on \\Movable Antennas}

\author{Jun Tang, Cunhua Pan,~\IEEEmembership{Senior Member,~IEEE}, Yang Zhang, Hong Ren,~\IEEEmembership{Member,~IEEE},\\and Kezhi Wang,~\IEEEmembership{Senior Member,~IEEE} 
\thanks{J. Tang is with the Faculty of Electrical Engineering
	and Computer Science, Ningbo University, Ningbo, China. (e-mail: tangjun001199@163.com). He will join Southeast University as a Master student. C. Pan, Y. Zhang and H. Ren are with National Mobile Communications Research Laboratory, Southeast University, Nanjing, China. (e-mail: cpan, 220230982, hren@seu.edu.cn). K. Wang is with the Department of Computer Science, Brunel University London, UB8 3PH Uxbridge, U.K. (e-mail: kezhi.wang@brunel.ac.uk).}
\thanks{\itshape{Corresponding authors:} Cunhua Pan and Hong Ren.}
}



\maketitle

\begin{abstract}
This paper considers a movable antenna (MA)-aided secure multiple-input multiple-output (MIMO) communication system consisting of a base station (BS), a legitimate information receiver (IR) and an eavesdropper (Eve), where the BS is equipped with MAs to enhance the system's physical layer security (PLS). Specifically, we aim to maximize the secrecy rate (SR) by jointly optimizing the transmit precoding (TPC) matrix, the artificial noise (AN) covariance matrix and the MAs' positions under the constraints of the maximum transmit power and the minimum distance between MAs. To solve this non-convex problem with highly coupled optimization variables, the block coordinate descent (BCD) method is applied to alternately update the variables. Specifically, we first reformulate the SR into a tractable form by utilizing the minimum mean square error (MMSE) method, and derive the optimal TPC matrix and the AN covariance matrix with fixed MAs' positions by applying the Lagrangian multiplier method in semi-closed forms. Then, the majorization-minimization (MM) algorithm is employed to iteratively optimize each MA's position while keeping others fixed. Finally, simulation results are provided to demonstrate the effectiveness of the proposed algorithms and the significant advantages of the MA-aided system over conventional fixed position antenna (FPA)-based system in enhancing system's security.
\end{abstract}


\section{Introduction}
\IEEEPARstart{W}{ith} the substantial increase of coverage and network heterogeneity, there have been severe concerns in terms of the security and privacy of the next-generation wireless communication\cite{9524814}. Over the last few decades, the cryptographic encryption/decryption has been regarded as the most important technology to the information security\cite{William2010}. However, to avoid the heavy computation and key management costs
caused by the encryption/decryption-based method, the physical layer security (PLS) technologies have attracted increasing concerns from both academia and industry recently\cite{8543573}.

In particular, the rapid increase in multi-antenna technologies has facilitated the application of secure beamforming/precoding techniques in PLS\cite{6582736}, and the optimal beamforming/precoding designs for the secrecy rate maximization (SRM) have been extensively studied in various scenarios\cite{5485016,5755208,6584932,9374557}. These studies have fully demonstrated the effectiveness of the secure beamforming/precoding techniques in improving the security performance of systems.

In addition, artificial noise (AN) is also widely applied in PLS related studies to interfere with the eavesdroppers (Eves) and improve the secrecy rate (SR)\cite{8543573}. Since the null space method of AN often fails to achieve the optimal security performance\cite{6826572,5524086}, the joint design of AN and the transmit beamforming/precoding in different scenarios has garnered more attention. Reference\cite{6587993} introduced the concept of generalized AN for the multiple-input single-output single eavesdropper (MISOSE) scenario, where AN is no longer restricted to the null space of the legitimate channel. Specifically, the authors reduced the SRM problem to a much simpler power allocation problem and solved the problem in an iterative manner. Furthermore, the SRM problem in the multiple-input single-output multiple eavesdroppers (MISOME) scenario was investigated in\cite{6482662} under perfect and imperfect channel state information (CSI), where the semi-definite programming (SDP) method was applied to jointly optimize the transmit beamforming matrix and AN. Simulation results demonstrated that the proposed AN-aided SRM design can achieve significant SR gains over the no-AN design and the isotropic AN design, particularly when there are more Eves. Moreover, the authors of \cite{6584932} demonstrated that AN can greatly improve the SR of multiple-input multiple-output (MIMO) systems in the multiple Eves scenario. Besides, the benefits derived from the joint optimization of AN and the transmit beamforming in the novel intelligent reflecting surface (IRS)-aided systems were also illustrated\cite{8972400,9024490,9201173,9293148}.
 
Although the effectiveness of secure beamforming/precoding and AN has been well-demonstrated, these technologies are often constrained by the transmit power, leading to performance limitations. In addition, the high correlation between the channels of the legitimate communication link and the eavesdropping link can even impede the ability of the aforementioned techniques to achieve high SR. Fortunately, a novel system framework aided by movable antennas (MAs) can be employed to reconstruct the channel, thereby breaking the bottleneck of security performance.

The concept of MA has been proposed recently to overcome the performance limitations of conventional fixed position antenna (FPA)-based communication systems. In MA-aided systems, the antennas are connected to the radio frequency (RF) chains via flexible cables to support the movement\cite{10318061}. In this way, the MAs can reconstruct the channel and improve the security performance of the systems due to a better utilization of the spatial degrees of freedom (DoFs). Specifically, the MAs are supposed to relocate to the positions that boost the channel condition for legitimate users while ensuring that Eves are situated in areas with deep fading, leading to an improved SR. Numerous studies have demonstrated the significant advantages of MA-aided systems over conventional FPA systems in signal power improvement, interference mitigation, flexible beamforming, spatial multiplexing and other aspects\cite{10286328}. The channel modeling for MA-aided systems was first investigated in\cite{10318061}. The authors proposed a field-response channel model for MA-aided systems under the far-field condition, and carried out performance analysis between the MA-aided and FPA systems under deterministic and statistical channels, which demonstrated the significant advantages of MAs in improving the received power and decreasing the outage probability. Furthermore, the results in\cite{10243545} showed that the joint design of the transmit covariance matrix and the positions of MAs can enhance the channel power and adjust the singular values of the channel matrix effectively, which can significantly improve the channel capacity of MIMO systems. Besides, the transmit power minimization and communication rate maximization problems were investigated in MA-aided multiuser scenarios\cite{10354003,xiao2023multiuser,cheng2023sumrate,qin2023antenna,wu2023movable}, where the antenna position optimization was employed to address these issues. Moreover, the authors of \cite{10278220} demonstrated that by optimizing the MAs' positions and the beamforming vector, the full array gain over the desired direction and the null steering over all undesired directions can be achieved simultaneously for MA arrays. In addition, the investigations in \cite{10382559,10414081,zhu2024performance} also substantiated the superiority of MA-aided systems in improving the performance of multi-beam forming, coordinated multi-point (CoMP) transmission and wideband communication. It is worth noting that the performance gain of MAs can only be achieved with the knowledge of accurate CSI. Thus, several works are also dedicated to estimating the channel precisely for MA-aided systems\cite{10236898,xiao2023channel}.

As previously mentioned, many studies have focused on the channel capacity improvement, multiuser communication, flexible beamforming and other topics in MA-aided systems. However, research on PLS in MA-aided systems is still in its infancy. Among the studies in terms of the MA-aided secure systems, reference\cite{cheng2023enabling} delved into a preliminary exploration of the MISOSE scenario. The authors aimed at minimizing the transmit power and maximizing the SR by jointly optimizing the MAs' positions and the transmit beamforming vector, and a gradient descent (GD)-based alternating optimization (AO) algorithm was proposed to solve the corresponding problems. Furthermore, the SRM problem for an MA array in the MISOME scenario was investigated in \cite{10416363}, where the MAs' positions and the transmit beamforming vector were jointly optimized. In summary, the existing works related to the PLS in MA-aided systems only studied the MISO scenario. However, with the advancement of antenna technology, devices equipped with multiple antennas are expected to be more prevalent in the future\cite{huang2014design}. Therefore, we investigate a more general MA-aided secure MIMO communication system in this paper. 

To the best of our knowledge, this is the first attempt to explore the PLS in MA-aided MIMO systems. The contributions of this paper are summarized as follows:
\begin{enumerate}
	\item We consider an MA-aided secure MIMO system where the base station (BS) is equipped with multiple MAs, while the information receiver (IR) and the Eve are equipped with multiple FPAs. Then, we formulate the SRM problem with respect to (w.r.t.) the transmit precoding (TPC) matrix, the AN covariance matrix and the MAs' positions, subject to the maximum transmit power and the minimum antenna spacing constraints. 
	\item To solve this non-convex problem with highly coupled variables, a method based on block coordinate descent (BCD) and majorization-minimization (MM) is proposed to alternately update the variables. Specifically, we first reformulate the objective function (OF) into a more tractable form by utilizing the minimum mean square error (MMSE) method and derive the optimal TPC matrix and the AN covariance matrix with fixed MAs' positions by applying the Lagrangian multiplier method in semi-closed forms. Then, the MM algorithm is employed to iteratively optimize the position of each MA while keeping others fixed.
	\item It is worth mentioning that many optimization variables in this paper can be derived in closed forms (or semi-closed forms), which makes the proposed algorithm computationally efficient.
	\item Simulation results validate the effectiveness of the proposed algorithm and demonstrate the significant advantages of the MA-aided system over conventional FPA system in security performance. Furthermore, we also observe that the increase in the size of transmit region, the number of antennas, the number of paths and the transmit power can enhance the security of the MA-aided system. Finally, the results also demonstrate that the accurate field-response information (FRI) is crucial to the MA-aided system, which will otherwise lead to severe degradation in the system's security performance.
\end{enumerate}

The remainder of this paper is organized as follows. In Section \ref{Sec_modeing}, we present the system and channel model of the MA-aided MIMO system and formulate the SRM problem. In Section \ref{Sec_algo}, we reformulate the original problem into a more tractable form, and develop a BCD-MM-based algorithm to tackle the problem. In Section \ref{Sec_simulation}, numerical simulation results are provided to demonstrate the advantages of MA-aided systems over FPA systems. Finally, we conclude this paper in Section \ref{Sec_conclusion}.

\emph{Notations}: Boldface lower case and
upper case letters denote vectors and matrices, respectively. ${\mathbb{C}}^{M\times N}$ and ${\mathbb{R}}^{M\times N}$ denote the space of $M\times N$ complex and real matrices, respectively. ${\mathbb{E}}\{\cdot\} $ denotes the expectation operation. ${\left\| {\bf{x}} \right\|_2}$ denotes the 2-norm of vector ${\bf{x}}$. ${\left\| {\bf{A}} \right\|_2}$ and ${\left\| {\bf{A}} \right\|_F}$ denote the spectral norm and the Frobenius norm of matrix ${\bf{A}}$, respectively. ${\rm{Tr}}\left( {\bf{A}} \right)$ and $\left| {\bf{A}} \right|$ denote the trace operation and the determinant of matrix ${\bf{A}}$, respectively. $\nabla_{\bf{x}} {f}\left( {\bf{x}} \right)$ and $\nabla^2_{\bf{x}} {f}\left( {\bf{x}} \right)$ denote the gradient vector and the Hessian matrix of the function $f$ w.r.t. the vector ${\bf{x}}$. ${\mathcal{CN}}\left({\bf{0}},{\bf{Z}}\right)$ denotes the circularly symmetric complex Gaussian (CSCG) distribution with zero mean and covariance matrix ${\bf{Z}}$. ${\rm{diag}}\left\{\cdot\right\}$ denotes the diagonalization operation. $\left(\cdot \right)^{T}$, $\left(\cdot \right)^{H}$ and $\left(\cdot \right)^{\dagger}$ denote the transpose, Hermitian transpose and pseudo-inverse operations, respectively. For a complex value $a$, ${\rm{Re}}\{a\}$ and $\angle a$ denote the real part and the angle of $a$, respectively. $\left[\cdot\right]^{+}$ denotes the projection onto the non-negative value, i.e., if $y=\left[x\right]^{+}$, then $y=\max\left\{0,x\right\}$.

\section{System Model and Problem Formulation}\label{Sec_modeing}
\subsection{System Model}
\begin{figure}[t]
	\centering
	\vspace*{12pt}
	\includegraphics[width=3.1in]{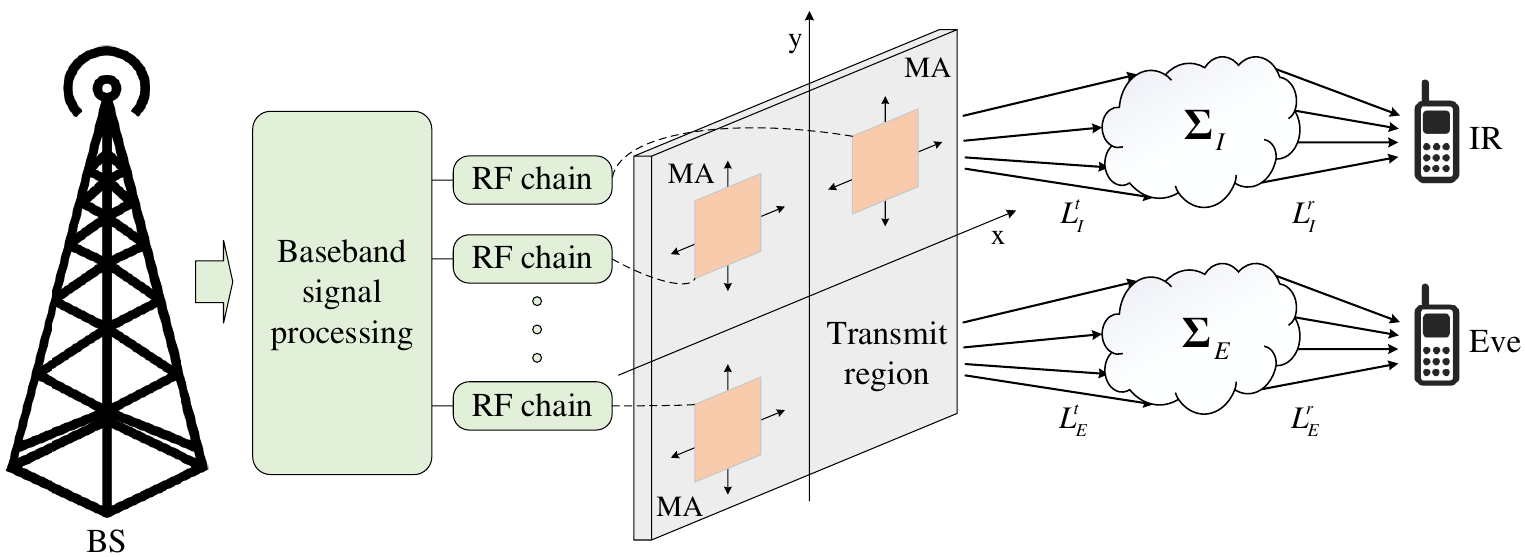}\vspace{-0.2cm}
	\caption{The MA-aided secure MIMO communication system.}
	\label{sym_model1}
\end{figure}

As shown in Fig. \ref{sym_model1}, the MA-aided secure MIMO communication system consists of a BS, a legitimate IR and an Eve, where the BS is equipped with $M\ge 2$ MAs, and the IR and the Eve are equipped with 
$N_I\ge 2$ and $N_E\ge 2$ FPAs, respectively. We denote ${\bf{V}}\in \mathbb{C}^{M\times d}$ as the TPC matrix, and ${\bf{s}}\in \mathbb{C}^{d\times 1}$, $d \le \min \left(M, N_I\right)$ as the data streams sent to the IR, respectively. Without loss of generality, we assume ${\bf{s}}\sim\mathcal{CN}\left({\bf{0}},{\bf{I}}_d\right)$.
Besides the information data, the AN vector is also employed to interfere with the Eve, which is denoted by ${\bf{z}}\sim\mathcal{CN}\left({\bf{0}},{\bf{R}}_z\right)$. Therefore, the transmit signal vector can be expressed as
\begin{equation}
	{\bf{x}} = {\bf{V}}{\bf{s}} + {\bf{z}}.
\end{equation}
We next denote the channel matrices from the BS to the IR and the Eve as ${\bf{H}}_I\in\mathbb{C}^{N_I\times M}$ and ${\bf{H}}_E\in\mathbb{C}^{N_E\times M}$, respectively. Then the signal received at the IR and the Eve are respectively given by
\begin{equation}
	{{\bf{y}}_I} = {{\bf{H}}_I}{\bf{x}} + {{\bf{n}}_I} = {{\bf{H}}_I}{\bf{Vs}} + {{\bf{H}}_I}{\bf{z}} + {{\bf{n}}_I},
\end{equation}
\begin{equation}
	{{\bf{y}}_E} = {{\bf{H}}_E}{\bf{x}} + {{\bf{n}}_E} = {{\bf{H}}_E}{\bf{Vs}} + {{\bf{H}}_E}{\bf{z}} + {{\bf{n}}_E},
\end{equation}
where ${\bf{n}}_I\sim \mathcal{CN}\left({\bf{0}},\sigma_I^2{\bf{I}}_{N_I}\right)$ and ${\bf{n}}_E\sim \mathcal{CN}\left({\bf{0}},\sigma_E^2{\bf{I}}_{N_E}\right)$ are the additive white Gaussian noise (AWGN) at the IR and the Eve, respectively. Hence, the rate of the IR and the Eve are respectively given by
\begin{equation}
	{R_I} = \log \left| {\bf{I}} + {{\bf{H}}_I}{\bf{V}}{{\bf{V}}^H}{\bf{H}}_I^H{\bf{J}}_I^{-1} \right|,
\end{equation}
\begin{equation}
	{R_E} = \log \left| {\bf{I}} + {{\bf{H}}_E}{\bf{V}}{{\bf{V}}^H}{\bf{H}}_E^H{\bf{J}}_E^{-1} \right|,
\end{equation}
where ${{\bf{J}}_I} \triangleq {{\bf{H}}_I}{\bf{R}}_z{\bf{H}}_I^H + \sigma _I^{2}{{\bf{I}}_{{N_I}}}$ and ${{\bf{J}}_E} \triangleq {{\bf{H}}_E}{\bf{R}}_z{\bf{H}}_E^H + \sigma _E^{2}{{\bf{I}}_{{N_E}}}$ are the interference-plus-noise covariance matrices at the IR and the Eve, respectively. Then, the SR is given by
\begin{equation}
		{R_S} = \left[{R_I} - {R_E}\right]^ +. 
\end{equation}

\subsection{Channel Model}\label{Channel model}
For an MA-aided communication system operating at a high-frequency band such as 28 GHz, the size of its transmit/receive region is much smaller than the signal propagation distance between the transmitter and the receiver, which makes the far-field condition easy to be satisfied\cite{10318061}. Thus, we adopt the planar wave model to construct the field-response channel model between the BS and the IR/Eve\cite{10318061}. Specifically, we assume that during the movement of the MAs, the signal for each path undergoes only phase variations, while the angle of departure (AoD), the angle of arrival (AoA) and the amplitude remain unchanged.
\begin{figure}[t]
	\centering
	\includegraphics[width=3.0in]{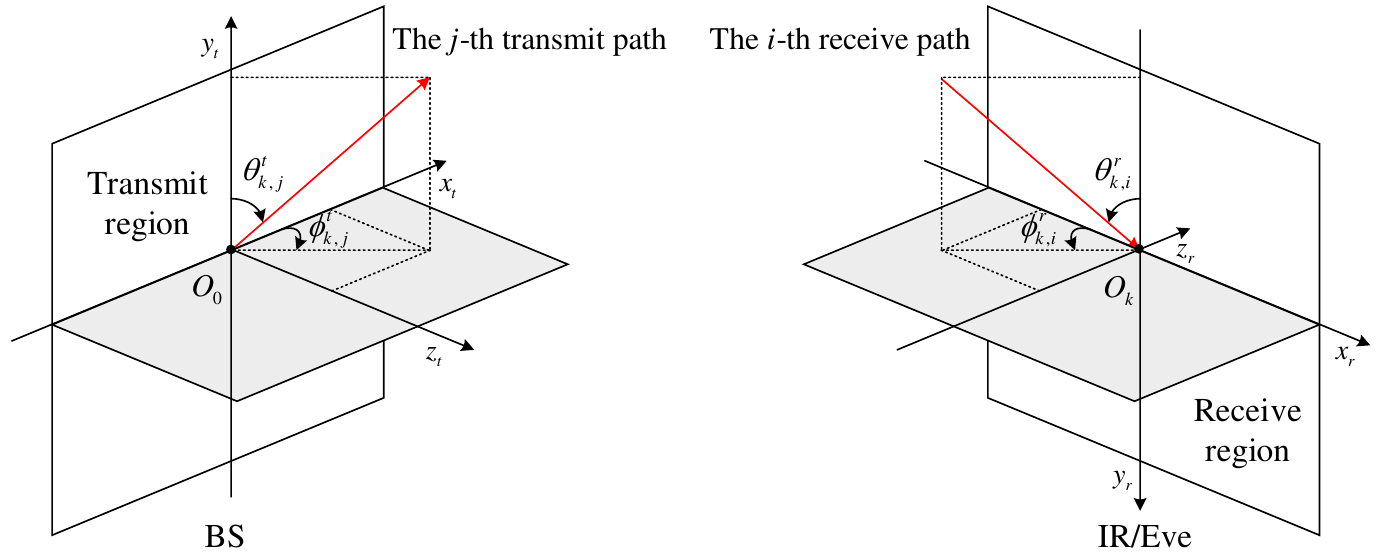}
	\caption{The far-field-response channel model.}
	\label{far_filed}
\end{figure}
We next denote ${\bf{T}}=\left[{\bf{t}}_1,\dots,{\bf{t}}_M\right]\in \mathbb{R}^{2\times M}$ as the positions of all $M$ MAs at the BS, where ${\bf{t}}_m=\left[x_m^t,y_m^t\right]^{T}$, $1\le m\le M$ represents the $m$-th MA's position. In addition, we denote $\mathcal{C}$ as the transmit region of the MAs. Without loss of generality, we assume $\mathcal{C}$ as an $A \times A$ square region, where $A$ denotes the length/width of the transmit region. Then, we define $k\in\left\{I,E\right\}$ as the set of the IR and the Eve for further manipulations, and the number of transmit and receive paths between the BS and the IR/Eve are denoted as $L_k^t$ and $L_k^r$, respectively. As shown in Fig. \ref{far_filed}, the transmit elevation and azimuth angles, along with the receive elevation and azimuth angles from the BS to the IR/Eve are denoted as $\theta_{k,j}^{t}\in \left[0,\pi\right]$, $\phi_{k,j}^{t}\in \left[0,\pi\right]$, $1\le j\le L_t^k$ and $\theta_{k,i}^{r}\in \left[0,\pi\right]$, $\phi_{k,i}^{r}\in \left[0,\pi\right]$, $1\le i\le L_k^r$, respectively. Moreover, we denote the position of the $n$-th antenna at the IR/Eve as ${\bf{v}}_{k,n}=\left[x_{k,n}^r,y_{k,n}^r\right]^T$, $1\le n\le N_k$. Then, the receive field-response vector (FRV) is given by
\begin{equation}
	{{\bf{f}}_k}\left({{\bf{v}}_{k,n}}\right) = {\left[{e^{j\frac{{2\pi }}{\lambda }\rho _{k,1}^r\left({{\bf{v}}_{k,n}}\right)}},\dots,{e^{j\frac{{2\pi }}{\lambda }\rho _{k,L_k^r}^r\left({{\bf{v}}_{k,n}}\right)}}\right]^T} \in \mathbb{C}^{L_k^r \times 1},
\end{equation}
where $\lambda$ denotes the wavelength, $\rho _{k,i}^r\left({{\bf{v}}_{k,n}}\right) = x_{k,n}^r\sin \theta _{k,i}^r\cos \phi _{k,i}^r + y_{k,n}^r\cos \theta _{k,i}^r$, $1\le i\le L_k^r$ denotes the difference of the signal propagation distance for the $i$-th receive path between the $n$-th antenna at the IR/Eve and their respective reference point $O_k$. As such, the receive field-response matrix (FRM) is given by
\begin{equation}
	{{\bf{F}}_k} = \left[ {{\bf{f}}_k}\left({{\bf{v}}_{k,1}}\right),\dots,{{\bf{f}}_k}\left({{\bf{v}}_{k,{N_k}}}\right)\right]  \in {\mathbb{C}^{L_k^r \times {N_k}}}.
\end{equation}
Similarly, the transmit FRV and FRM are given by
\begin{equation}
	{{\bf{g}}_k}\left({{\bf{t}}_m}\right) = {\left[ {e^{j\frac{{2\pi }}{\lambda }\rho _{k,1}^t\left({{\bf{t}}_m}\right)}},\dots,{e^{j\frac{{2\pi }}{\lambda }\rho _{k,L_k^t}^t\left({{\bf{t}}_m}\right)}}\right] ^T} \in {\mathbb{C}^{L_k^t \times 1}},
\end{equation}
\begin{equation}
	{{\bf{G}}_k} = \left[{{\bf{g}}_k}\left({{\bf{t}}_1}\right),\dots,{{\bf{g}}_k}\left({{\bf{t}}_M}\right)\right]\in {\mathbb{C}^{L_k^t \times M}},
\end{equation}
respectively, where $\rho _{k,j}^t\left({{\bf{t}}_m}\right) = x_m^t\sin \theta _{k,j}^t\cos \phi _{k,j}^t + y_m^t\cos \theta _{k,j}^t$, $1\le j\le L_k^t$ denotes the difference of the signal propagation distance for the $j$-th transmit path between the $m$-th MA and the BS's reference point $O_0$.

In addition, we define the path-response matrix (PRM) ${\bf{\Sigma}}_k\in \mathbb{C}^{L_k^r\times L_k^t}$ as the response coefficients for all paths from the BS's reference point $O_0$ to the IR/Eve's reference point $O_k$. Then, the channel matrix ${\bf{H}}_k\in \mathbb{C}^{N_k\times M}$ from the BS to the IR/Eve is given by
\begin{equation}\label{channel matrix}
	{{\bf{H}}_k} = {\bf{F}}_k^H{\bf{\Sigma}}_k{{\bf{G}}_{k}}.
\end{equation}
Since the IR and the Eve are both equipped with FPAs, in the following part of this paper, ${\bf{F}}_k$ and ${\bf{\Sigma}}_k$ are assumed to be constant, while ${\bf{G}}_k$ is regarded as a function of ${\bf{T}}$.

\subsection{Problem Formulation}
In this paper, we aim to maximize the SR by jointly optimizing the TPC matrix ${\bf{V}}$, the AN covariance matrix ${\bf{R}}_z$ and the MAs' positions ${\bf T}$ subject to constraints of the maximum transmit power at the BS and the minimum distance between MAs. Firstly, let ${{\bf{R}}_z} = {{\bf{V}}_E}{\bf{V}}_E^H$ and we can formulate the SRM problem as
\begin{subequations}\label{initial_p}
	\begin{align}
		\mathop {\max }\limits_{{\bf{V}},{\bf{V}}_E,{\bf{T}}}
		\quad& R_S \label{initial_p_of}\\
		\textrm{s.t.} \qquad& {\rm{Tr}}\left({\bf{V}}{\bf{V}}^H+{{\bf{V}}_E}{\bf{V}}_E^H\right)\le P_{\max},\\
		\quad& {{\bf{t}}_m} \in {\mathcal C},1\le m\le M,\\
		\quad& \left\| {{{\bf{t}}_m} - {\bf{t}}_{m^\prime} } \right\|_2 \ge D,m \ne m^\prime \label{initial_p_constraint},
	\end{align}
\end{subequations}
where $P_{\max}$ is the maximum transmit power, and $D$ is the minimum distance between MAs to avoid the coupling effect. Notably, the above problem is hard to solve since the OF is non-concave and the constraint (\ref{initial_p_constraint}) is non-convex. Moreover, the optimization variables are highly coupled, making Problem (\ref{initial_p}) more intractable.

\section{Proposed BCD-MM-based Algorithm}\label{Sec_algo}
In this section, we first reformulate the OF of Problem (\ref{initial_p}) into a more tractable form. Then, the BCD-MM-based algorithm is proposed to alternately optimize the TPC matrix ${\bf {V}}$, the decomposition of AN covariance matrix ${\bf{V}}_E$, and the MAs' positions ${\bf{T}}$.
\subsection{Problem Reformulation}
Firstly, the SR can be reformulated as
\begin{equation}
	\begin{aligned}
		R_S
		\\=&\log \left|{\bf{I}}_{N_I}+{\bf{H}}_I {\bf{V}} {\bf{V}}^H {\bf{H}}_I^H\left({\bf{H}}_I {\bf{V}}_E {\bf{V}}_E^H {\bf{H}}_I^H+\sigma_I^2 {\bf{I}}_{N_I}\right)^{-1}\right|\\
		&+\log \left|{\bf{H}}_E {\bf{V}}_E {\bf{V}}_E^H {\bf{H}}_E^H+\sigma_E^2 {\bf{I}}_{N_E}\right|\\
		&-\log \left|{\bf{H}}_E {\bf{V}}_E {\bf{V}}_E^H {\bf{H}}_E^H+\sigma_E^2 {\bf{I}}_{N_E}+{\bf{H}}_E {\bf{V}} {\bf{V}}^H {\bf{H}}_E^H\right| \\
		=&\underbrace{\log \left|{\bf{I}}_{N_I}+{\bf{H}}_I {\bf{V}} {\bf{V}}^H {\bf{H}}_I^H\left({\bf{H}}_I {\bf{V}}_E {\bf{V}}_E^H {\bf{H}}_I^H+\sigma_I^2 {\bf{I}}_{N_I}\right)^{-1}\right|}_{f_1} \\
		&+\underbrace{\log \left|{\bf{I}}_{N_E}+{\bf{H}}_E {\bf{V}}_E {\bf{V}}_E^H {\bf{H}}_E^H\left(\sigma_E^2 {\bf{I}}_{N_E}\right)^{-1}\right|}_{f_2} \\
		&\underbrace{-\log \left|{\bf{I}}_{N_E}+\sigma_E^{-2} {\bf{H}}_E\left({\bf{V}} {\bf{V}}^H+{\bf{V}}_E {\bf{V}}_E^H\right) {\bf{H}}_E^H\right|}_{f_3},
	\end{aligned}
\end{equation}
which is still non-concave and difficult to tackle. Therefore, we reformulate the expressions of $f_1$, $f_2$ and $f_3$ into more tractable forms. In terms of $f_1$ and $f_2$, we adopt the MMSE method\cite{shi2015secure} by transforming the expression of rate to another equivalent but more tractable form via introducing auxiliary matrices. Specifically, for $f_1$, we first introduce ${\bf U}_I\in \mathbb{C}^{N_I\times d}$ as the linear decoding matrix to estimate the data streams, i.e., $\hat{{\bf{s}}}={\bf U}_I^H{\bf{y}}_I$. Then, the mean square error (MSE) matrix of estimation is given by
\begin{equation}
	\begin{aligned}
		{{\bf{E}}_I}\left( {{{\bf{U}}_I},{\bf{V}},{{\bf{V}}_E}} \right) \triangleq &\ \mathbb{E}\left[ {\left(\hat {\bf{s}} - {\bf{s}}\right){{\left(\hat {\bf{s}} - {\bf{s}}\right)}^H}} \right] \\
		=& \left( {{\bf{U}}_I^H{{\bf{H}}_I}{\bf{V}} - {{\bf{I}}_d}} \right){\left( {{\bf{U}}_I^H{{\bf{H}}_I}{\bf{V}} - {{\bf{I}}_d}} \right)^H} \\
		&+ {\bf{U}}_I^H\left( {{{\bf{H}}_I}{{\bf{V}}_E}{{\bf{V}}_E^H}{\bf{H}}_I^H + \sigma _I^2{{\bf{I}}_{{N_I}}}} \right){{\bf{U}}_I}.
	\end{aligned}
\end{equation}
In addition, we introduce an auxiliary matrix ${\bf W}_I\in \mathbb{C}^{d\times d}$, ${\bf{W}}_I \succeq {\bf{0}}$. According to Lemma 4.1 in \cite{shi2015secure}, $f_1$ can be reformulated as
\begin{equation}
	\begin{aligned}
		f_1 =&\mathop {\max }\limits_{{\bf{U}}_I, {\bf{W}}_I} h_1\left({\bf{U}}_I, {\bf{V}}, {\bf{V}}_E, {\bf{W}}_I\right) \\
		\triangleq &\mathop {\max }\limits_{{\bf{U}}_I, {\bf{W}}_I} \log \left|{\bf{W}}_I\right|-{\rm{Tr}}\left({\bf{W}}_I {\bf{E}}_I\left({\bf{U}}_I, {\bf{V}}, {\bf{V}}_E\right)\right)+d.
	\end{aligned}
\end{equation}
Note that $h_1\left({\bf{U}}_I, {\bf{V}}, {\bf{V}}_E, {\bf{W}}_I\right)$ is concave w.r.t. ${\bf{U}}_I$ and ${\bf{W}}_I$ with other variables fixed. Thus, by considering the first-order optimality condition, the optimal solutions of ${\bf{U}}_I$ and ${\bf{W}}_I$ can be derived as
\begin{equation}\label{U_I}
	\begin{aligned}
		{\bf{U}}_I^ \star =& \arg \max _{{{\bf{U}}_I}} h_1\left( {{{\bf{U}}_I},{\bf{V}},{{\bf{V}}_E},{{\bf{W}}_I}} \right) \\
		=& {\left( {{{\bf{H}}_I}{{\bf{V}}_E}{\bf{V}}_E^H{\bf{H}}_I^H + \sigma _I^2{{\bf{I}}_{{N_I}}} + {{\bf{H}}_I}{\bf{V}}{{\bf{V}}^H}{\bf{H}}_I^H} \right)^{ - 1}}{{\bf{H}}_I}{\bf{V}},
	\end{aligned}
\end{equation}
\begin{equation}\label{W_I}
	\begin{aligned}
		{\bf{W}}_I^ \star  = \arg \max _{{{\bf{W}}_I}} h_1\left( {{{\bf{U}}_I},{\bf{V}},{{\bf{V}}_E},{{\bf{W}}_I}} \right) 
		= {\left[ {{{\bf{E}}_I}\left( {{\bf{U}}_I^ \star ,{\bf{V}},{{\bf{V}}_E}} \right)} \right]^{ - 1}},
	\end{aligned}
\end{equation}
respectively. Similarly, for $f_2$, we first introduce ${\bf U}_E\in \mathbb{C}^{N_E\times M}$ as the decoding matrix, and the MSE matrix of estimation is given by
\begin{equation}
	\begin{aligned}
		{{\bf{E}}_E}\left( {{{\bf{U}}_E},{{\bf{V}}_E}} \right) = &\left( {{\bf{U}}_E^H{{\bf{H}}_E}{{\bf{V}}_E} - {{\bf{I}}_M}} \right){\left( {{\bf{U}}_E^H{{\bf{H}}_E}{{\bf{V}}_E} - {{\bf{I}}_M}} \right)^H} \\
		&+ \sigma _E^2{\bf{U}}_E^H{{\bf{U}}_E}.
	\end{aligned}
\end{equation}
By introducing an auxiliary matrix ${{\bf{W}}_E}\in \mathbb{C}^{M\times M}$, ${\bf{W}}_E \succeq {\bf{0}}$, $f_2$ can be reformulated as
\begin{equation}
	\begin{aligned}
		{f_2} = &\mathop {\max }\limits_{{{\bf{U}}_E},{{\bf{W}}_E}}h_2\left( {{{\bf{U}}_E},{{\bf{V}}_E},{{\bf{W}}_E}} \right) \\
		\triangleq& \mathop {\max }\limits_{{{\bf{U}}_E},{{\bf{W}}_E}}\log \left| {{{\bf{W}}_E}} \right| - {\mathop{{\rm {Tr}}}\nolimits} \left( {{{\bf{W}}_E}{{\bf{E}}_E}\left( {{{\bf{U}}_E},{{\bf{V}}_E}} \right)} \right) + M,
	\end{aligned}
\end{equation}
and the optimal solutions of ${\bf{U}}_E$ and ${\bf{W}}_E$ can be derived as
\begin{equation}\label{U_E}
	\begin{aligned}
		{\bf{U}}_E^ \star  = & \arg \max _{{{\bf{U}}_E}}h_2\left( {{{\bf{U}}_E},{{\bf{V}}_E},{{\bf{W}}_E}} \right) \\
		=& {\left( {\sigma _E^2{{\bf{I}}_{{N_E}}} + {{\bf{H}}_E}{{\bf{V}}_E}{\bf{V}}_E^H{\bf{H}}_E^H} \right)^{ - 1}}{{\bf{H}}_E}{{\bf{V}}_E},
	\end{aligned}
\end{equation}
\begin{equation}\label{W_E}
	\begin{aligned}
		{\bf{W}}_E^ \star  = \arg \max _{{{\bf{W}}_E}} h_2\left( {{{\bf{U}}_E},{{\bf{V}}_E},{{\bf{W}}_E}} \right) 
		= {\left[ {{{\bf{E}}_E}\left( {{\bf{U}}_E^ \star ,{{\bf{V}}_E}} \right)} \right]^{ - 1}},
	\end{aligned}
\end{equation}
respectively. Next, in terms of the expression of $f_3$, we introduce the following lemma.

\itshape \textbf{Lemma 1:}  \upshape Let ${\bf{E}}\in \mathbb{C}^{N \times N}$ be any matrix such that ${\bf{E}}\succeq {\bf{0}}$. Consider the function $f\left({\bf{W}}\right)=-{\rm {Tr}}\left({\bf{WE}}\right)+\log |{\bf{W}}|+N$. Then, we have
\begin{equation}\label{lemm2_eq}
	\log \left| {{{\bf{E}}^{ - 1}}} \right| = {\max _{{\bf{W}\succeq{\bf{0}}} }}f({\bf{W}}),
\end{equation}
and the optimal solution is given by ${{\bf{W}}^ \star } = {{\bf{E}}^{ - 1}}$.

\itshape \textbf{Proof:} \upshape Please refer to\cite{5962875}.\hfill $\blacksquare$

According to Lemma 1, we first define ${{\bf{E}}_X}\left( {{\bf{V}},{{\bf{V}}_E}} \right) \triangleq {{\bf{I}}_{{N_E}}} + \sigma _E^{ - 2}{{\bf{H}}_E}\left( {{\bf{V}}{{\bf{V}}^H} + {{\bf{V}}_E}{\bf{V}}_E^H} \right){\bf{H}}_E^H$ and introduce an auxiliary matrix ${\bf{W}}_X\in \mathbb{C}^{N_E \times N_E}$, ${{\bf W}_X}\succeq {\bf {0}}$, then $f_3$ can be reformulated as
\begin{equation}
	\begin{aligned}
		f_3 =& \mathop {\max }\limits_{{\bf{W}}_X} h_3\left({\bf{V}},{\bf{V}}_E,{\bf{W}}_X\right) \\
		\triangleq & \mathop {\max }\limits_{{\bf{W}}_X}\log \left|{\bf{W}}_X\right| - {\mathop{{\rm {Tr}}}\nolimits} \left( {{{\bf{W}}_X}{{\bf{E}}_X}\left( {{\bf{V}},{{\bf{V}}_E}} \right)} \right) + N_E,
	\end{aligned}
\end{equation}
where the optimal solution is given by 
\begin{equation}\label{W_X}
	{\bf{W}}_X^ \star  = {\left[{{\bf{E}}_X}\left({\bf{V}},{{\bf{V}}_E}\right)\right]^{ - 1}}.
\end{equation}

After the above steps, the SR can be rewritten as
\begin{equation}
	{R_S} = {\max _{{{\bf{U}}_I},{{\bf{W}}_I},{{\bf{U}}_E},{{\bf{W}}_E},{{\bf{W}}_X}}}{F}\left({\bf{K}}\right),
\end{equation}
where ${\bf{K}} \triangleq \left[ {{{\bf{U}}_I},{{\bf{W}}_I},{{\bf{U}}_E},{{\bf{W}}_E},{{\bf{W}}_X},{\bf{V}},{{\bf{V}}_E},{\bf{T}}} \right]$ and ${F}\left({\bf{K}}\right) \triangleq {h_1}\left( {{{\bf{U}}_I},{\bf{V}},{{\bf{V}}_E},{{\bf{W}}_I}} \right) + {h_2}\left( {{{\bf{U}}_E},{{\bf{V}}_E},{{\bf{W}}_E}} \right) + {h_3}\left( {{\bf{V}},{{\bf{V}}_E},{{\bf{W}}_X}} \right)$, respectively. Then, Problem (\ref{initial_p}) can be reformulated as
\begin{subequations}\label{bcd_p}
	\begin{align}
		\mathop {\max }\limits_{{\bf{K}}}\quad& F\left({\bf{K}}\right) \label{bcd_p_OF}\\
		\textrm {s.t.}\quad&  {\rm {Tr}}\left({\bf{V}} {\bf{V}}^{H}+{\bf{V}}_E {\bf{V}}_E^H\right) \leq P_{\max }, \\
		& {\bf{t}}_m \in \mathcal{C}, 1\le m \le M,\\
		& \left\|{\bf{t}}_m-{\bf{t}}_{m^{\prime}}\right\|_2 \geq D, m \neq m^{\prime},\\
		& {\bf W}_I\succeq {\bf{0}},{\bf W}_E\succeq {\bf{0}},{\bf W}_X\succeq {\bf{0}}.
	\end{align}
\end{subequations}
Note that the OF (\ref{bcd_p_OF}) is concave w.r.t. $\left\lbrace {{\bf{U}}_I},{{\bf{W}}_I},{{\bf{U}}_E},{{\bf{W}}_E},{{\bf{W}}_X}\right\rbrace$ with fixed $\left\lbrace {\bf{V}},{{\bf{V}}_E},{\bf{T}}\right\rbrace $, and concave w.r.t. $\left\lbrace {\bf{V}},{{\bf{V}}_E}\right\rbrace $ with fixed $\left\lbrace {{\bf{U}}_I},{{\bf{W}}_I},{{\bf{U}}_E},{{\bf{W}}_E},{{\bf{W}}_X},{\bf{T}}\right\rbrace$, respectively. Thus, Problem (\ref{bcd_p}) is more tractable than Problem (\ref{initial_p}) and the BCD algorithm can be applied to optimize each variable alternately. Specifically, we first update  ${{\bf{U}}_I},{{\bf{W}}_I},{{\bf{U}}_E},{{\bf{W}}_E},{{\bf{W}}_X}$ in closed forms with fixed ${\bf{V}},{{\bf{V}}_E},{\bf{T}}$. Then, with
given ${{\bf{U}}_I},{{\bf{W}}_I},{{\bf{U}}_E},{{\bf{W}}_E},{{\bf{W}}_X}$, we remove the constant terms and update ${\bf{V}},{{\bf{V}}_E},{\bf{T}}$ by solving the following subproblem
\begin{subequations}
	\begin{align}
		\mathop {\min}\limits_{{\bf{V}},{{\bf{V}}_E},{\bf{T}}}
		\quad& \xi\left({\bf{V}},{{\bf{V}}_E},{\bf{T}}\right) \\
		\textrm{s.t.}\qquad & {\rm{Tr}}\left({\bf{V}} {\bf{V}}^{H}+{\bf{V}}_E {\bf{V}}_E^H\right) \leq P_{\max }, \\
		& {\bf{t}}_m \in \mathcal{C}, 1 \le m \le M, \\
		& \left\|{\bf{t}}_m-{\bf{t}}_{m^{\prime}}\right\|_2 \geq D, m \neq m^{\prime},
	\end{align}
\end{subequations}
where
\begin{equation}
	\begin{aligned}
		\xi&\left({\bf{V}},{{\bf{V}}_E},{\bf{T}}\right)\\ \triangleq& -2{\rm {Re}}\left\lbrace {\rm {Tr}}\left( {{{\bf{W}}_I}{{\bf{V}}^H}{\bf{H}}_I^H{{\bf{U}}_I}} \right)\right\rbrace +{\rm {Tr}} \left( 	{{{\bf{V}}^H}{{\bf{H}}_V}{\bf{V}}} \right)\\&
		-2{\rm {Re}}\left\lbrace {\rm {Tr}}\left( {{{\bf{W}}_E}{\bf{V}}_E^H{{\bf{H}}_E^H}{{\bf{U}}_E}}\right)\right\rbrace +{\rm{Tr}} \left( {{\bf{V}}_E^H{{\bf{H}}_{VE}}{{\bf{V}}_E}} \right),
	\end{aligned}
\end{equation}
\begin{equation}
	{{\bf{H}}_V} = {\bf{H}}_I^H{{\bf{U}}_I}{{\bf{W}}_I}{\bf{U}}_I^H{{\bf{H}}_I} + \sigma _E^{ - 2}{\bf{H}}_E^H{{\bf{W}}_X}{{\bf{H}}_E},
\end{equation}
\begin{equation}
	\begin{aligned}
		{{\bf{H}}_{VE}} =& \  {\bf{H}}_I^H{{\bf{U}}_I}{{\bf{W}}_I}{\bf{U}}_I^H{{\bf{H}}_I} + {\bf{H}}_E^H{{\bf{U}}_E}{{\bf{W}}_E}{\bf{U}}_E^H{{\bf{H}}_E} \\
		&\ +\sigma _E^{ - 2}{\bf{H}}_E^H{{\bf{W}}_X}{{\bf{H}}_E},
	\end{aligned}
\end{equation}
respectively, and it is readily verified that both ${\bf{H}}_V$ and ${\bf{H}}_{VE}$ are positive semi-definite matrices.

\subsection{Optimize ${\bf{V}}$ and ${\bf{V}}_E$}
In this subsection, we aim to optimize ${\bf{V}}$ and ${\bf{V}}_E$ with fixed ${\bf{T}}$, and the subproblem is given by
\begin{subequations}\label{QCQP}
	\begin{align}
		\mathop {\min }\limits_{{\bf{V}},{{\bf{V}}_E}} \quad& \xi\left({\bf{V}},{{\bf{V}}_E}\right)\\
		\textrm {s.t.} \quad &{\rm{Tr}}\left({\bf{V}} {\bf{V}}^{H}+{\bf{V}}_E {\bf{V}}_E^H\right) \leq P_{\max }.
	\end{align}
\end{subequations}
Note that the above problem is a convex quadratically constrained quadratic program (QCQP) problem, and a low complexity algorithm based on the Lagrangian multiplier method was proposed in\cite{9201173} to obtain the optimal solutions of ${\bf{V}}$ and ${\bf{V}}_E$ in semi-closed forms as
\begin{equation}\label{optimal_V}
	{\bf{V}^{\star}}=\left({\bf{H}}_V+\lambda{\bf{I}}\right)^{\dagger}\left({\bf{H}}_I^H{\bf{U}}_I{\bf{W}}_I\right),
\end{equation}
\begin{equation}\label{optimal_VE}
	{{\bf{V}}_E^{\star}}=\left({\bf{H}}_{VE}+\lambda{\bf{I}}\right)^{\dagger}\left({\bf{H}}_E^H{\bf{U}}_E{\bf{W}}_E\right),
\end{equation}
respectively, where $\lambda$ denotes the Lagrangian multiplier and can be derived by utilizing the bisection search method to satisfy the complementary slackness condition. We recommend the readers to refer to\cite{9201173} for more details.  

\subsection{Optimize ${\bf{T}}$}\label{Optimize_T}
In this subsection, we aim to optimize the MAs' positions ${\bf{T}}$ with fixed ${\bf{V}}$ and ${\bf{V}}_E$, and the subproblem is given by
\begin{subequations}\label{optimal_T}
	\begin{align}
		\mathop {\min }\limits_{{\bf{T}}} \quad& \xi\left({\bf{T}}\right)\\
		\textrm{s.t.}\quad & {\bf{t}}_m \in \mathcal{C}, 1\le m \le M, \\
		& \left\|{\bf{t}}_m-{\bf{t}}_{m^{\prime}}\right\|_2 \geq D, m \neq m^{\prime}.
	\end{align}
\end{subequations}
Since the variation of ${\bf{T}}$ will only affect the transmit FRM ${\bf{G}}_k$ according to Section \ref{Channel model}, we then reformulate ${\xi}\left({\bf{T}}\right)$ into a more tractable form w.r.t. ${\bf{G}}_k$ in the following part of this subsection.

Firstly, recalling the definitions of channel matrices ${\bf{H}}_I$ and ${\bf{H}}_E$ in (\ref{channel matrix}), we have
\begin{subequations}\label{other_items}
	\begin{align}
		2{\rm {Re}}\left\lbrace {\rm {Tr}}\left( {{{\bf{W}}_I}{{\bf{V}}^H}{\bf{H}}_I^H{{\bf{U}}_I}} \right)\right\rbrace =2{\rm {Re}}\left\{{\rm{Tr}}\left({\bf{A}}_I{\bf{G}}_I^H \right)\right\},\\
		2{\rm {Re}}\left\lbrace {\rm {Tr}}\left( {{{\bf{W}}_E}{\bf{V}}_E^H{{\bf{H}}_E^H}{{\bf{U}}_E}}\right)\right\rbrace =2{\rm {Re}}\left\{{\rm{Tr}}\left({\bf{A}}_E{\bf{G}}_E^H \right)\right\},
	\end{align}		
\end{subequations} 
where 
\begin{equation}
	{{\bf{A}}_I} = {{\bf{\Sigma }}_I^H}{\bf{F}}_I{\bf{U}}_I{{\bf{W}}_I}{\bf{V}}^H,
\end{equation}
\begin{equation}
	{{\bf{A}}_E} = {{\bf{\Sigma }}_E^H}{\bf{F}}_E{\bf{U}}_E{{\bf{W}}_E}{{\bf{V}}_E^H},
\end{equation}
respectively, and the summation of the second and the fourth terms of ${\xi}\left({\bf{T}}\right)$ is given by
\begin{equation}\label{sum_of_3_6_fix}
	\begin{aligned}
		{\rm{Tr}}&\left({\bf{V}}^H {\bf{H}}_V {\bf{V}}\right)+{\rm{Tr}}\left({\bf{V}}_E^H {\bf{H}}_{V E} {\bf{V}}_E\right)\\
		=&\ {\rm{Tr}}\left({\bf{G}}_I {\bf{V}}_X {\bf{G}}_I^H {\bf{C}}_I\right) + {\rm {Tr}}\left({\bf{G}}_E {\bf{V}}_X {\bf{G}}_E^H {\bf{C}}_X\right) \\
		&\ + {\rm {Tr}}\left({\bf{G}}_E {\bf{V}}_E {\bf{V}}_E^H {\bf{G}}_E^H {\bf{C}}_E\right), 
	\end{aligned}
\end{equation}
where ${\bf{V}}_X={\bf{V}} {\bf{V}}^H+{\bf{V}}_E {\bf{V}}_E^H$, ${{\bf{C}}_I} = {\bf{\Sigma }}_I^H{{\bf{F}}_I}{\bf{U}}_I {\bf{W}}_I {\bf{U}}_I^H{\bf{F}}_I^H{{\bf{\Sigma }}_I}$, ${{\bf{C}}_X} = \sigma _E^{ - 2}{\bf{\Sigma }}_E^H{{\bf{F}}_E}{{\bf{W}}_X}{\bf{F}}_E^H{{\bf{\Sigma }}_E}$, ${\bf{C}}_E={\bf{\Sigma}}_E^H {\bf{F}}_E {{\bf{U}}_E {\bf{W}}_E {\bf{U}}_E^H} {\bf{F}}_E^H {\bf{\Sigma}}_E$, respectively.

Next, we manage to optimize the $m$-th MA's position ${\bf{t}}_m$ while keeping others, i.e., $\left\lbrace {{{\bf{t}}_{m'}},m' \ne m} \right\rbrace _{m' = 1}^M$ fixed. Specifically, we decouple the $m$-th MA's FRV ${\bf{g}}_k({\bf{t}}_m)$ from ${\xi}\left({\bf{T}}\right)$ as

\begin{equation}\label{sum_all_items}
	\begin{aligned}
		\xi\left({\bf{T}}\right)
		=&\ 
		{\bf{g}}_I^H\left({{\bf{t}}_m}\right){{\bf{B}}_{I,m}}{{\bf{g}}_I}\left({{\bf{t}}_m}\right) +{\bf{g}}_E^H\left({{\bf{t}}_m}\right){{\bf{B}}_{E,m}}{{\bf{g}}_E}\left({{\bf{t}}_m}\right)\\ &+ 2{\rm{Re}}\left\lbrace  {\bf{g}}_I^H\left({{\bf{t}}_m}\right){{\bf{d}}_{I,m}}\right\rbrace   + 2{\rm{Re}}\left\lbrace  {\bf{g}}_E^H\left({{\bf{t}}_m}\right){{\bf{d}}_{E,m}}\right\rbrace \\
		&+\textrm{const} ,
	\end{aligned}
\end{equation}
where
\begin{equation}
	{\bf{B}}_{I, m}={\bf{q}}_{X, m}^H {\bf{q}}_{X, m} {\bf{C}}_I,
\end{equation}
\begin{equation}
	{{\bf{B}}_{E,m}} = {{\bf{q}}_{X,m}^H}{\bf{q}}_{X,m}{{\bf{C}}_X} + {{\bf{q}}_{E,m}^H}{\bf{q}}_{E,m}{{\bf{C}}_E},
\end{equation}
\begin{equation}
	{{\bf{d}}_{I,m}} = {{\bf{C}}_I} \left( \sum\limits_{i = 1,i \ne m}^M {{\bf{g}}_I}\left( {{\bf{t}}_i} \right) {{\bf{q}}_{X,i}^H}\right) {\bf{q}}_{X,m}- {\bf{a}}_{I,m},
\end{equation}
\begin{equation}
	\begin{aligned}
		{{\bf{d}}_{E,m}} =& \ {{\bf{C}}_X}\left( \sum\limits_{i = 1,i \ne m}^M {{\bf{g}}_E}\left( {{\bf{t}}_i}\right) {{\bf{q}}_{X,i}^H}\right){\bf{q}}_{X,m} \\
		&+ {{\bf{C}}_E}\left( \sum\limits_{i = 1,i \ne m}^M {{\bf{g}}_E}\left( {{\bf{t}}_i} \right) {{\bf{q}}_{E,i}^H}\right) {\bf{q}}_{E,m} - {\bf{a}}_{E,m},
	\end{aligned}
\end{equation}
respectively. The detailed proof of (\ref{sum_all_items}) and the definitions of ${\bf{q}}_{X,i}\in \mathbb{C}^{M\times 1}$, ${\bf{q}}_{E,i} \in \mathbb{C}^{M\times 1}$, ${{\bf{a}}_{I,m}}\in \mathbb{C}^{L_I^t\times 1}$ and ${{\bf{a}}_{E,m}}\in \mathbb{C}^{L_E^t\times 1}$ are given in Appendix A. Then, by removing the constant terms from ${\bf{\xi}}\left({\bf{T}}\right)$, Problem (\ref{optimal_T}) can be reformulated as
\begin{subequations}\label{before_mm}
	\begin{align}
		\mathop{\min}\limits_{{\bf{t}}_m}\quad &{\bf{g}}_I^H\left({{\bf{t}}_m}\right){{\bf{B}}_{I,m}}{{\bf{g}}_I}\left({{\bf{t}}_m}\right) +{\bf{g}}_E^H\left({{\bf{t}}_m}\right){{\bf{B}}_{E,m}}{{\bf{g}}_E}\left({{\bf{t}}_m}\right)\label{objective function}\nonumber\\
	    &+ 2{\rm{Re}}\left\lbrace  {\bf{g}}_I^H\left({{\bf{t}}_m}\right){{\bf{d}}_{I,m}}\right\rbrace  + 2{\rm{Re}}\left\lbrace  {\bf{g}}_E^H\left({{\bf{t}}_m}\right){{\bf{d}}_{E,m}}\right\rbrace \\
	    \textrm {s.t.} \quad &{\bf{t}}_m \in \mathcal{C}, \\
	    &\left\|{\bf{t}}_m-{\bf{t}}_{m^{\prime}}\right\|_2 \geq D, m \neq m^{\prime}\label{beform_non-convex-constraint}.
	\end{align}
\end{subequations}
Since Problem (\ref{before_mm}) is still non-convex due to the OF and the constraint (\ref{beform_non-convex-constraint}), the classical iterative approach, i.e., MM algorithm\cite{7547360} is employed to tackle this problem. Specifically, at the $i$-th iteration, we denote the solution of Problem (\ref{before_mm}) as ${\bf{t}}_m^i$, and the OF value as $f\left({\bf{t}}_m^i\right)$. Then, at the $\left(i+1\right)$-th iteration, we need to introduce $\varepsilon \left({{\bf{t}}_m}|{\bf{t}}_m^i\right)$ which satisfies the following three conditions as a surrogate function of $f\left({\bf{t}}_m^i\right)$.
\begin{enumerate}
\item $\varepsilon ({\bf{t}}_m^i|{\bf{t}}_m^i)=f({\bf{t}}_m^i)$,
\item $\nabla_{{\bf{t}}_m}\varepsilon({{\bf{t}}_m}|{\bf{t}}_m^i)|_{{\bf{t}}_m={\bf{t}}_m^i}=\nabla_{{\bf{t}}_m}f({{\bf{t}}_m}) |_{{{\bf{t}}_m}={{\bf{t}}_m^i}}$,
\item $\varepsilon({{\bf{t}}_m}|{\bf{t}}_m^i)\ge f({{\bf{t}}}_m)$.
\end{enumerate}
The first and the second conditions mean that the value and the gradient of the surrogate function at ${\bf{t}}_m^i$ should be equal to the original OF. The third condition means that the surrogate function should be an upper bound of the original OF. Then, we can solve the approximate subproblem by replacing the original OF with the surrogate function $\varepsilon \left({{\bf{t}}_m}|{\bf{t}}_m^i\right)$ at the $\left(i+1\right)$-th iteration. In this way, the sequence of the solutions obtained in each iteration will result in a monotonically decreasing optimal value $\left\{f({{\bf{t}}_m^i}),i = 1,2,\dots \right\}$ and finally converge.

However, note that the OF (\ref{objective function}) is non-concave w.r.t. ${\bf{g}}_k\left({{\bf{t}}_m}\right)$, so the surrogate function cannot be obtained by the first-order Taylor expansion at ${\bf{g}}_k({\bf{t}}_m^i)$ of the original OF. Alternatively, we introduce the following lemma to construct the surrogate function.

\itshape \textbf{Lemma 2:}  \upshape For any given solution ${\bf{t}}_m^i$ at the $i$-th iteration and for any feasible ${\bf{t}}_m$, we have
\begin{equation}
	\begin{aligned}
		{\bf{g}}_k^H&\left({\bf{t}}_m\right) {\bf{B}}_{k, m} {\bf{g}}_k\left({\bf{t}}_m\right)  \\
		\leq\ & {\bf{g}}_k^H\left({\bf{t}}_m\right) {\bf{\Phi}}_k {\bf{g}}_k\left({\bf{t}}_m\right)\\
		&-2{\rm{Re}}\left\lbrace {\bf{g}}_k^H\left({\bf{t}}_m\right)\left({\bf{\Phi}}_k-{\bf{B}}_{k, m}\right) {\bf{g}}_k\left({\bf{t}}_m^i\right)\right\rbrace \\
		&+{\bf{g}}_k^H\left({\bf{t}}_m^i\right)\left({\bf{\Phi}}_k-{\bf{B}}_{k, m}\right) {\bf{g}}_k\left({\bf{t}}_m^i\right) \\
		\triangleq\  & \mu_k\left({\bf{t}}_m \mid {\bf{t}}_m^i\right),
	\end{aligned}
\end{equation}
where ${\bf{\Phi}}_k = \lambda _{\max}^k{\bf{I}}_{L_k^t}$ and $\lambda _{\max}^k$ is the maximum eigenvalue of ${\bf{B}}_{k,m}$, $k\in\left\{I,E\right\}$.

\itshape \textbf{Proof:}  \upshape Please refer to\cite{7093191}. \hfill $\blacksquare$

According to Lemma 2, we can construct the surrogate function as
\begin{equation}
	\begin{aligned}
		\tau \left({{\bf{t}}_m}|{\bf{t}}_m^i\right) \triangleq \ & {\mu _I}\left({{\bf{t}}_m}|{\bf{t}}_m^i\right) + 2{\rm{Re}}\left\lbrace  {\bf{g}}_I^H\left({{\bf{t}}_m}\right){{\bf{d}}_{I,m}}\right\rbrace \\
		&+ {\mu _E}\left({{\bf{t}}_m}|{\bf{t}}_m^i\right)+ 2{\rm{Re}}\left\lbrace  {\bf{g}}_E^H\left({{\bf{t}}_m}\right){{\bf{d}}_{E,m}}\right\rbrace,  
	\end{aligned}
\end{equation}
which satisfies the three conditions as a surrogate function. In addition, since ${\bf{g}}_k^H\left({\bf{t}}_m\right){\bf{g}}_k\left({\bf{t}}_m\right)=L_k^t$, we have ${\bf{g}}_k^H\left({\bf{t}}_m\right){\bf{\Phi}}_k{\bf{g}}_k\left({\bf t}_m\right)=L_k^t\lambda _{\max}^k$ as a constant, and $\tau \left({{\bf{t}}_m}|{\bf{t}}_m^i\right)$ can be recast as
\begin{equation}
	\begin{aligned}
		\tau \left({{\bf{t}}_m}|{\bf{t}}_m^i\right) =\ & 2{\rm{Re}}\left\lbrace {\bf{g}}_I^H\left({\bf{t}}_m\right) {\boldsymbol{\eta}}_I^i+{\bf{g}}_E^H\left({\bf{t}}_m\right) {\boldsymbol{\eta}}_E^i\right\rbrace  + \textrm{const},
	\end{aligned}
\end{equation}
where
\begin{equation}\label{Xi_I}
	{\boldsymbol{\eta }}_I^i = {{\bf{d}}_{I,m}} - \left({{\bf{\Phi }}_I} - {{\bf{B}}_{I,m}}\right){{\bf{g}}_I}\left({\bf{t}}_m^i\right),
\end{equation}
\begin{equation}\label{Xi_E}
	{\boldsymbol{\eta }}_E^i = {{\bf{d}}_{E,m}} - \left({{\bf{\Phi }}_E} - {{\bf{B}}_{E,m}}\right){{\bf{g}}_E}\left({\bf{t}}_m^i\right),
\end{equation}
respectively. Then, we define $\gamma \left({\bf{t}}_m\right)\triangleq \tau \left({{\bf{t}}_m}|{\bf{t}}_m^i\right)$ for further manipulations. Note that $\gamma \left({\bf{t}}_m\right)$ is still non-concave w.r.t. ${{\bf{t}}_m}$, and its upper bound cannot be obtained by the first-order Taylor expansion. Thus, we utilize the second-order Taylor expansion to construct the surrogate function of $\gamma \left({\bf{t}}_m\right)$\cite{10243545}. Specially, we introduce ${\delta _m} > 0$, which satisfies ${\delta _m}{{\bf{I}}_2}\succeq {\nabla ^2}\gamma \left({\bf{t}}_m\right)$\cite{Magnus1995MatrixDC}, then we have 
\begin{equation}\label{gamma_tm}
	\begin{aligned}
		\gamma\left({\bf{t}}_m\right) \leq &\  \gamma\left({\bf{t}}_m^i\right)+\nabla \gamma\left({\bf{t}}_m^i\right)^T\left({\bf{t}}_m-{\bf{t}}_m^i\right)\\
		&+\frac{\delta_m}{2}\left({\bf{t}}_m-{\bf{t}}_m^i\right)^T\left({\bf{t}}_m-{\bf{t}}_m^i\right) \\
		=&\ \frac{\delta_m}{2} {\bf{t}}_m^T {\bf{t}}_m+\left(\nabla \gamma\left({\bf{t}}_m^i\right)-\delta_m {\bf{t}}_m^i\right)^T {\bf{t}}_m\\
		&+\underbrace{\gamma\left({\bf{t}}_m^i\right)+\frac{\delta_m}{2}\left({\bf{t}}_m^i\right)^T {\bf{t}}_m^i-\nabla \gamma\left({\bf{t}}_m^i\right)^T {\bf{t}}_m^i}_{\textrm {const }}\\
		\triangleq&\ \varepsilon \left({\bf{t}}_m|{\bf{t}}_m^i\right),
	\end{aligned}
\end{equation}
where the gradient vector $\nabla \gamma \left( {{{\bf{t}}_m}} \right)$ and the Hessian matrix ${\nabla ^2}\gamma \left( {{{\bf{t}}_m}} \right)$ w.r.t. ${\bf{t}}_m$ are given in Appendix B, and the construction of $\delta _m$ is given in Appendix C.
It is readily verified that $\varepsilon \left({\bf{t}}_m|{\bf{t}}_m^i\right)$ satisfies the three conditions as a surrogate function. Therefore, by dropping the constant terms in $\varepsilon \left({\bf{t}}_m|{\bf{t}}_m^i\right)$ at the $\left(i+1\right)$-th iteration, Problem (\ref{before_mm}) can be approximated as
\begin{subequations}\label{QP_no_convex}
	\begin{align}
		\mathop{\min}\limits_{{\bf{t}}_m}\quad &\frac{{{\delta _m}}}{2}{\bf{t}}_m^T{{\bf{t}}_m} + {\left( {\nabla \gamma \left( {{\bf{t}}_m^i} \right) - {\delta _m}{\bf{t}}_m^i} \right)^T}{{\bf{t}}_m}\label{QP_no_convex_OF}\\
		\textrm{s.t.}\quad &{\bf{t}}_m \in \mathcal{C}, \label{box_constraint}\\
		&\left\|{\bf{t}}_m-{\bf{t}}_{m^{\prime}}\right\|_2 \geq D, m \neq m^{\prime},\label{no_convex}
	\end{align}
\end{subequations}
which is still non-convex owing to the constraint (\ref{no_convex}). Note that the OF (\ref{QP_no_convex_OF}) is a convex quadratic function w.r.t. ${\bf{t}}_m$, we can first neglect the constraints (\ref{box_constraint}) and (\ref{no_convex}), and then obtain the optimal ${\bf{t}}_m$ in a closed form\cite{10243545}, i.e., 
\begin{equation}\label{tm_closed_form}
	{\bf{t}}_m^{\star}={\bf{t}}_m^i-\frac{1}{\delta_m}\nabla\gamma\left({\bf{t}}_m^i\right).
\end{equation}
However, if ${\bf{t}}_m^{\star}$ does not satisfy the constraints (\ref{box_constraint}) and (\ref{no_convex}), we have to turn back to tackle Problem (\ref{QP_no_convex}) to obtain a feasible solution. Specifically, we can relax the constraint (\ref{no_convex}) by applying the first-order Taylor expansion of $\left\|{\bf{t}}_m-{\bf{t}}_{m^{\prime}}\right\|_2$ as
\begin{equation}\label{2-norm_taylor}
	\begin{aligned}
	\left\|{\bf{t}}_m-{\bf{t}}_{m^{\prime}}\right\|_2 \geq &\left\|{\bf{t}}_m^i-{\bf{t}}_{m^{\prime}}\right\|_2\\
	&+\left(\left.\nabla\left\|{\bf{t}}_m-{\bf{t}}_{m^{\prime}}\right\|_2\right|_{{\bf{t}}_m={\bf{t}}_m^i}\right)^T\left({\bf{t}}_m-{\bf{t}}_m^i\right) \\
	=&\frac{1}{\left\|{\bf{t}}_m^i-{\bf{t}}_{m^{\prime}}\right\|_2}\left({\bf{t}}_m^i-{\bf{t}}_{m^{\prime}}\right)^T\left({\bf{t}}_m-{\bf{t}}_{m^{\prime}}\right).
	\end{aligned}
\end{equation}
By substituting the left hand side of (\ref{no_convex}) with (\ref{2-norm_taylor}), Problem (\ref{QP_no_convex}) can be recast as
\begin{subequations}\label{QP_convex}
	\begin{align}
		\mathop{\min}\limits_{{\bf{t}}_m}\quad &\frac{{{\delta _m}}}{2}{\bf{t}}_m^T{{\bf{t}}_m} + {\left( {\nabla \gamma \left( {{\bf{t}}_m^i} \right) - {\delta _m}{\bf{t}}_m^i} \right)^T}{{\bf{t}}_m}\\
		\textrm{s.t.}\quad &{\bf{t}}_m \in \mathcal{C}, \\
		&\frac{1}{{{{\left\| {{\bf{t}}_m^i - {{\bf{t}}_{m'}}} \right\|}_2}}}{\left( {{\bf{t}}_m^i - {{\bf{t}}_{m'}}} \right)^T}\left( {{{\bf{t}}_m} - {{\bf{t}}_{m'}}} \right) \ge D,m\ne m^{\prime},
	\end{align}
\end{subequations}
which is a convex quadratic programming (QP) problem and can be solved by quadprog\cite{turlach2007quadprog} via the interior-point method\cite{ben2001lectures}. Based on the above discussions, the detailed procedures of the MM algorithm are summarized in Algorithm \ref{algo_1}. When the algorithm converges, we can obtain the $m$-th MA's optimal position. 
\begin{algorithm}
	\caption{MM Algorithm for Solving Problem (\ref{before_mm})}\label{MM_Algo}
	\begin{algorithmic}[1]\label{algo_1}
		\STATE Initialize the iteration index $i=1$, and the convergence threshold $\varepsilon_1$. Input the feasible solution ${\bf{t}}_m^1$ and calculate the OF value of Problem (\ref{before_mm}) as $f\left({\bf{t}}_m^1\right)$;
		\STATE Calculate ${\bf{\Phi}}_I$ and ${\bf{\Phi}}_E$, according to Lemma 2;
		\STATE Calculate ${\boldsymbol{\eta}}_I^i$ and ${\boldsymbol{\eta}}_E^i$ via (\ref{Xi_I}) and (\ref{Xi_E}), respectively;
		\STATE Calculate $\nabla \gamma \left( {{{\bf{t}}_m^i}} \right)$, ${\nabla ^2}\gamma \left( {{{\bf{t}}_m^i}} \right)$ and $\delta_m$ via (\ref{first_order_grad}), (\ref{second_order_hessen}) and (\ref{value of delta_m}), respectively;
		\STATE Obtain ${\bf{t}}_m^{\star}$ via (\ref{tm_closed_form});
		\IF{${\bf{t}}_m^{\star}$ satisfice the constraints (\ref{box_constraint}) and (\ref{no_convex})} 
		\STATE Set ${\bf{t}}_m^{i+1}\leftarrow{\bf{t}}_m^{\star}$;
		\ELSE
		\STATE Obtain a feasible ${\bf{t}}_m^{i+1}$ via solving Problem (\ref{QP_convex});
		\ENDIF 
		\STATE Calculate $f\left({\bf{t}}_m^{i+1}\right)$, if $\left|f\left({\bf{t}}_m^{i+1}\right)-f\left({\bf{t}}_m^{i}\right)\right|/\left|f\left({\bf{t}}_m^{i+1}\right)\right|\le\varepsilon_1$ holds, terminate; Otherwise, set $i\leftarrow i+1$ and go to step 3.
	\end{algorithmic}
\end{algorithm}

According to the above discussions, it is readily verified that the OF value of (\ref{before_mm}) will monotonically decrease and finally converge since we have
\begin{equation}
	f\left({\bf{t}}_m^{i+1}\right)\overset{\left(a\right)}{\le} \varepsilon({{\bf{t}}_m^{i+1}}|{\bf{t}}_m^i)\overset{\left(b\right)}{\le}
	\varepsilon({{\bf{t}}_m^{i}}|{\bf{t}}_m^i)\overset{\left(c\right)}{=}f\left({\bf{t}}_m^{i}\right),
\end{equation}
where $\left(a\right)$ comes from the third condition of the surrogate function, $\left(b\right)$ comes from the fact that ${\bf{t}}_m^{i+1}$ is the optimal solution at each iteration and $\left(c\right)$ comes from the first condition of the surrogate function.

In the following, we analyze the computational complexity of Algorithm \ref{algo_1}: Firstly, at step 2, we need to calculate the maximum eigenvalues of ${\bf{B}}_{I,m}$ and ${\bf{B}}_{E,m}$, and the complexities are given by $\mathcal{O}\left({L_I^t}^3\right)$ and $\mathcal{O}\left({L_E^t}^3\right)$, respectively. Then, at step 3, we need to calculate ${\boldsymbol{\eta}}_I^i$ and ${\boldsymbol{\eta}}_E^i$, which results in the complexity of $\mathcal{O}\left({L_I^t}^2\right)+\mathcal{O}\left({L_E^t}^2\right)$. Next, from step 4 to step 5, the complexities required to calculate the value of $\nabla\gamma\left({\bf{t}}_m^i\right)$, $\nabla^2\gamma\left({\bf{t}}_m^i\right)$, $\delta_m$ and ${\bf{t}}_m^{\star}$ are  
$\mathcal{O}\left({L_I^t}+{L_E^t}\right)$, $\mathcal{O}\left({L_I^t}+{L_E^t}\right)$, $\mathcal{O}\left(1\right)$ and $\mathcal{O}\left(1\right)$, respectively. Besides, as mentioned above, if ${\bf{t}}_m^{\star}$ derived by (\ref{tm_closed_form}) is infeasible, we need to update ${\bf{t}}_m^{i+1}$ via solving Problem (\ref{QP_convex}), incurring the complexity of $\mathcal{O}\left(M^{1.5}\ln\left(1/\beta\right)\right)$, where $\beta$ denotes the accuracy for the interior-point method\cite{ben2001lectures}. Let $N_1$ and $N_2$ denote the maximum iteration times to repeat steps 3-10 and the maximum iteration times to solve Problem (\ref{QP_convex}), respectively. Then, the total complexity of the algorithm is given by $\mathcal{O}\left({L_I^t}^3+{L_E^t}^3+N_1\left({L_I^t}^2+{L_E^t}^2\right)+N_2M^{1.5}\ln\left(1/\beta\right)\right)$.

\subsection{Overall Algorithm}
Based on the above discussions, the detailed procedures of the overall BCD-MM-based algorithm are summarized in Algorithm \ref{algo_2}. It is readily verified that the OF value of Problem (\ref{bcd_p}) monotonically increases in each step of Algorithm \ref{algo_2}. In addition, the OF value has an upper bound due to the power constraint. Thus, the convergence of the proposed algorithm is guaranteed.

\begin{algorithm}
	\caption{Overall BCD-MM-based Algorithm for Solving the SRM Problem}\label{Overall_BCD_Algorithm}
	\begin{algorithmic}[1]\label{algo_2}
		\STATE Initialize the iteration index $n=1$, the maximum number of iterations $n_{\max}$, and the convergence threshold $\varepsilon_2$. Input the feasible solution ${\bf{V}}^{\left(1\right)}$, ${\bf{V}}_E^{\left(1\right)}$ and ${\bf{T}}^{\left(1\right)}$ and calculate the OF value of Problem (\ref{bcd_p}) as $F\left({\bf{V}}^{\left(1\right)},{\bf{V}}_E^{\left(1\right)},{\bf{T}}^{\left(1\right)}\right)$;
		
		\STATE Given ${\bf{V}}^{\left(n\right)}$, ${\bf{V}}_E^{\left(n\right)}$ and ${\bf{T}}^{\left(n\right)}$, calculate the optimal decoding matrices ${\bf{U}}_I^{\left(n\right)}$ and ${\bf{U}}_E^{\left(n\right)}$ via (\ref{U_I}) and (\ref{U_E}), respectively;
		
		\STATE Given ${\bf{V}}^{\left(n\right)}$, ${\bf{V}}_E^{\left(n\right)}$, ${\bf{T}}^{\left(n\right)}$, ${\bf{U}}_I^{\left(n\right)}$ and ${\bf{U}}_E^{\left(n\right)}$, calculate the optimal auxiliary matrices ${\bf{W}}_I^{\left(n\right)}$, ${\bf{W}}_E^{\left(n\right)}$ and ${\bf{W}}_X^{\left(n\right)}$ via (\ref{W_I}), (\ref{W_E}) and (\ref{W_X}), respectively;
		
		\STATE Given ${\bf{U}}_I^{\left(n\right)}$, ${\bf{U}}_E^{\left(n\right)}$, ${\bf{W}}_I^{\left(n\right)}$, ${\bf{W}}_E^{\left(n\right)}$, ${\bf{W}}_X^{\left(n\right)}$, ${\bf{T}}^{\left(n\right)}$, update the TPC matrix ${\bf{V}}^{\left(n+1\right)}$ and the decomposition of AN covariance matrix ${\bf{V}}_E^{\left(n+1\right)}$ via (\ref{optimal_V}) and (\ref{optimal_VE}), respectively;
		
		\FOR{$m=1$ to $M$}  
		\STATE Update ${\bf{t}}_m^{\left(n+1\right)}$ via Algorithm \ref{algo_1};
		\ENDFOR 
		
		\STATE Calculate $F\left({\bf{V}}^{\left(n+1\right)},{\bf{V}}_E^{\left(n+1\right)},{\bf{T}}^{\left(n+1\right)}\right)$, if $n>n_{\max}$ or
		$\frac{\left|F\left({\bf{V}}^{\left(n+1\right)},{\bf{V}}_E^{\left(n+1\right)},{\bf{T}}^{\left(n+1\right)}\right)-F\left({\bf{V}}^{\left(n\right)},{\bf{V}}_E^{\left(n\right)},{\bf{T}}^{\left(n\right)}\right)\right|}{\left|F\left({\bf{V}}^{\left(n+1\right)},{\bf{V}}_E^{\left(n+1\right)},{\bf{T}}^{\left(n+1\right)}\right)\right|}\le\varepsilon_2$ holds, terminate; Otherwise, set $n\leftarrow n+1$ and go to step 2.
	\end{algorithmic}
\end{algorithm}

The computational complexity of Algorithm \ref{algo_2} is analyzed as follows: Firstly, at step 2, we need to calculate the optimal decoding matrices ${\bf{U}}_I^{\left(n\right)}$ and ${\bf{U}}_E^{\left(n\right)}$, and the complexities are given by $\mathcal{O}\left(N_I^3\right)$ and $\mathcal{O}\left(N_E^3\right)$, respectively. Then, at step 3, the complexities of calculating the optimal auxiliary matrices ${\bf{W}}_I^{\left(n\right)}$, ${\bf{W}}_E^{\left(n\right)}$ and ${\bf{W}}_X^{\left(n\right)}$ are given by $\mathcal{O}\left(d^3\right)$, $\mathcal{O}\left(M^3\right)$ and $\mathcal{O}\left(N_E^3\right)$, respectively. Next, at step 4, we need to update ${\bf{V}}^{\left(n+1\right)}$ and ${\bf{V}}_E^{\left(n+1\right)}$, resulting in the complexity of  $\mathcal{O}\left(\max\left\lbrace 2M^3,2M^2N_E\right\rbrace \right)$ with the assumption of $M>N_I(\text{or }N_E)>d$\cite{9201173}. Finally, from step 5 to step 7, the MM algorithm is applied to update the MAs' positions ${\bf{T}}^{\left(n+1\right)}$, incurring the complexity of  $\mathcal{C}_{MM}=\mathcal{O}\Big(M\left({L_I^t}^3+{L_E^t}^3+N_1\left({L_I^t}^2+{L_E^t}^2\right)\right)+N_2M^{2.5}\ln\left(1/\beta\right)\Big)$. Thus, the overall complexity of the algorithm can be expressed as  $\mathcal{O}\left(\max\left\lbrace 2M^3,2M^2N_E,\mathcal{C}_{MM}\right\rbrace \right)$.

\section{Simulation Results}\label{Sec_simulation}
In this section, simulation results are provided to
verify the effectiveness of the proposed algorithm and the significant advantages of the MA-aided system over conventional FPA system in improving the system’s security performance. We first provide the simulation setup and then compare the proposed MA scheme with several baseline schemes. Finally, we investigate the influence of the imperfect FRI on the MA-aided system's security performance.
\subsection{Simulation Setup}
Unless otherwise stated, the simulation parameters are set as follows: The BS is equipped with $M=4$ MAs, the IR and the Eve are equipped with $N_I=N_E=4$ FPA-based uniform planar arrays (UPAs), and the number of data streams is given by $d\triangleq\min\left(M,N_I\right)$. Besides, the distance between the IR/Eve and the BS is assumed to be a random variable obeying the uniform distribution, i.e., $d_k\sim\mathcal{U}\left[d_{\min},d_{\max}\right],k\in\left\{I,E\right\}$, where $d_{\min}=20$ m and $d_{\max}=100$ m denote the nearest and the farthest distance, respectively. The elevation and azimuth AoAs of the IR/Eve, along with the elevation and azimuth AoDs of the BS are assumed to be independent identically distributed (i.i.d.) variables following the uniform distribution, i.e., $\theta_{k,i}^{r}\sim\mathcal{U}\left[0,\pi\right]$, $\phi_{k,i}^{r}\sim\mathcal{U}\left[0,\pi\right]$, $1\le i\le L_k^r$ and $\theta_{k,j}^{t}\sim\mathcal{U}\left[0,\pi\right]$,
$\phi_{k,j}^{t}\sim\mathcal{U}\left[0,\pi\right]$, $1\le j\le L_t^k$, respectively. Furthermore, the geometric channel model is adopted to describe the channel between the BS and the IR/Eve. Specifically, the number of transmit and receive paths are identical, i.e., $L_k^t=L_k^r=L_k$, and we assume $L_k=L=6$. In this way, the PRM ${\bf{\Sigma}}_k\in \mathbb{C}^{L_k^r\times L_k^t}$ is represented as a diagonal matrix with the element $\sigma_{ll}^k\sim \mathcal{CN}(0,g_0d_k^{-\alpha}/L)$, $1\le l\le L$, where $g_0=-40$ dB is the average channel gain at the reference distance $d_0=1$ m, $\alpha=2.8$ is the path-loss exponent. In addition, we set the maximum transmit power of $P_{\max}=10$ dBm, the noise power of $\sigma_I^2=\sigma_E^2=-80$ dBm, the minimum distance between MAs of $D=\frac{\lambda}{2}$, the wavelength of $\lambda = 0.01$ m, and the transmit region of $\mathcal{C}=\left[ -\frac{A}{2}, \frac{A}{2} \right] \times \left[ -\frac{A}{2}, \frac{A}{2} \right]$, where $A=4\lambda$ is the size of the transmit region. The convergence thresholds for Algorithm \ref{algo_1} and Algorithm \ref{algo_2} are set to $\varepsilon_1=10^{-7}$ and $\varepsilon_2=10^{-5}$, respectively. All the results are averaged over 400 independent channel realizations. 

\subsection{Convergence Behaviour of the Proposed Algorithms}
\begin{figure}[t]
	\centering
	\includegraphics[width=3.25in]{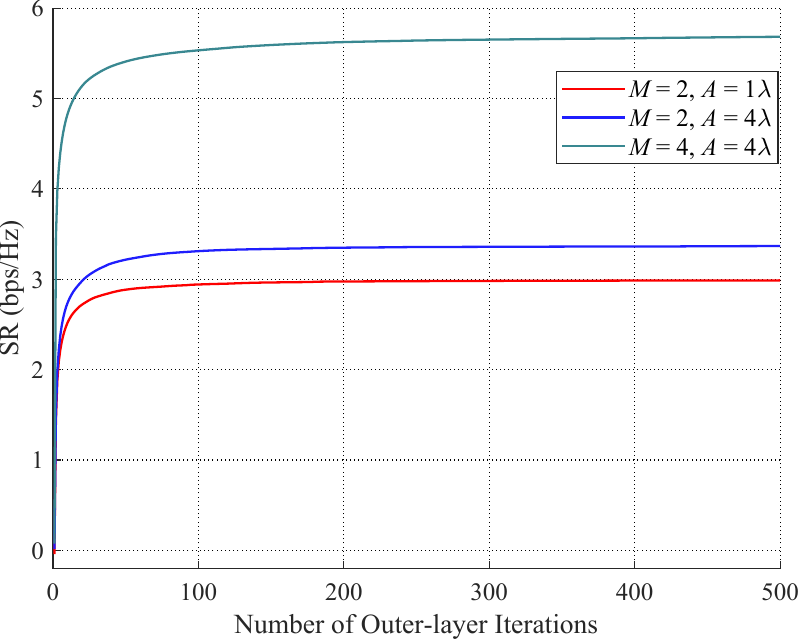}\vspace{-0.3cm}
	\caption{Convergence behaviour of the BCD algorithm.}
	\label{BCD_algorithm}
\end{figure}

Firstly, the convergence behaviour of the proposed overall BCD-MM-based algorithm is illustrated in Fig. \ref{BCD_algorithm}. We can observe that for different numbers of MAs and the sizes of transmit region, the achieved SR monotonically increases and finally converges to a stable value within about 200 iterations. Specifically, when $M=4$ and $A=4\lambda$, the achieved SR increases from 0.08 bps/Hz to 5.68 bps/Hz, which verifies the effectiveness of the proposed algorithm in enhancing the security of the MA-aided system. Moreover, we can observe that the increase in $M$ and $A$ leads to a higher achieved SR due to a better utilization of the spatial DoFs.

\begin{figure}[t]
	\centering
	\includegraphics[width=3.25in]{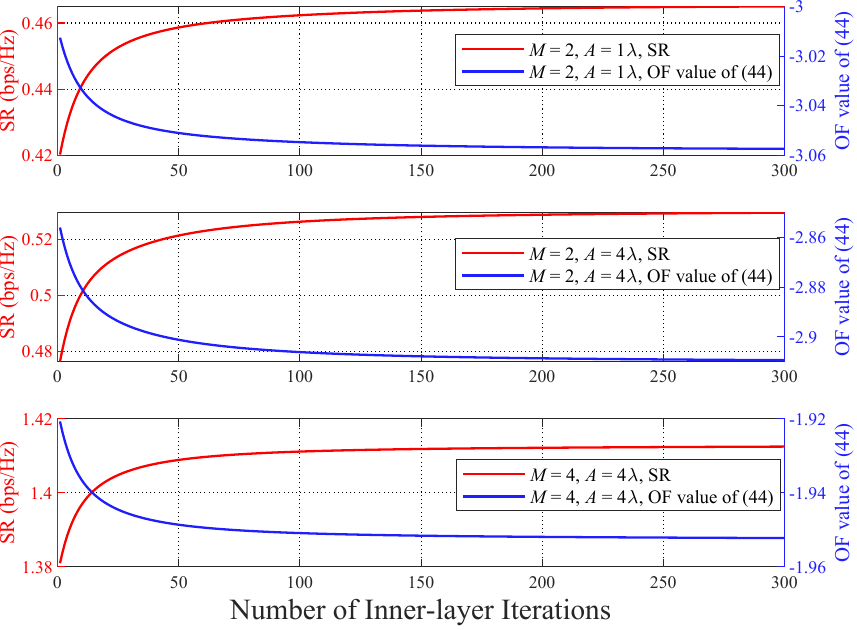}\vspace{-0.3cm}
	\caption{Convergence behaviour of the MM algorithm.}
	\label{MM_algorithm}
\end{figure}

Besides, the convergence behaviour of the MM algorithm is also investigated. As shown in Fig. \ref{MM_algorithm}, for different  numbers of MAs and the sizes of transmit region, the achieved SR monotonically increases, while the OF value of Problem (\ref{before_mm}) monotonically decreases, which is consistent with our previous discussions on the convergence of the MM algorithm in Section \ref{Optimize_T}.
\begin{figure}[t]
	\centering
	\includegraphics[width=3.25in]{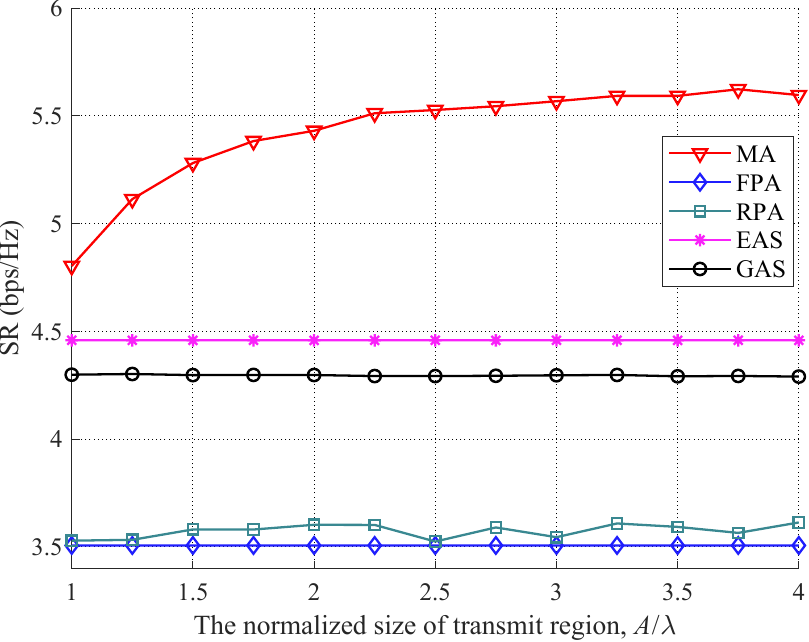}\vspace{-0.3cm}
	\caption{SR versus the size of transmit region.}
	\label{vs_A}
\end{figure}
\begin{figure}[t]
	\centering
	\includegraphics[width=3.25in]{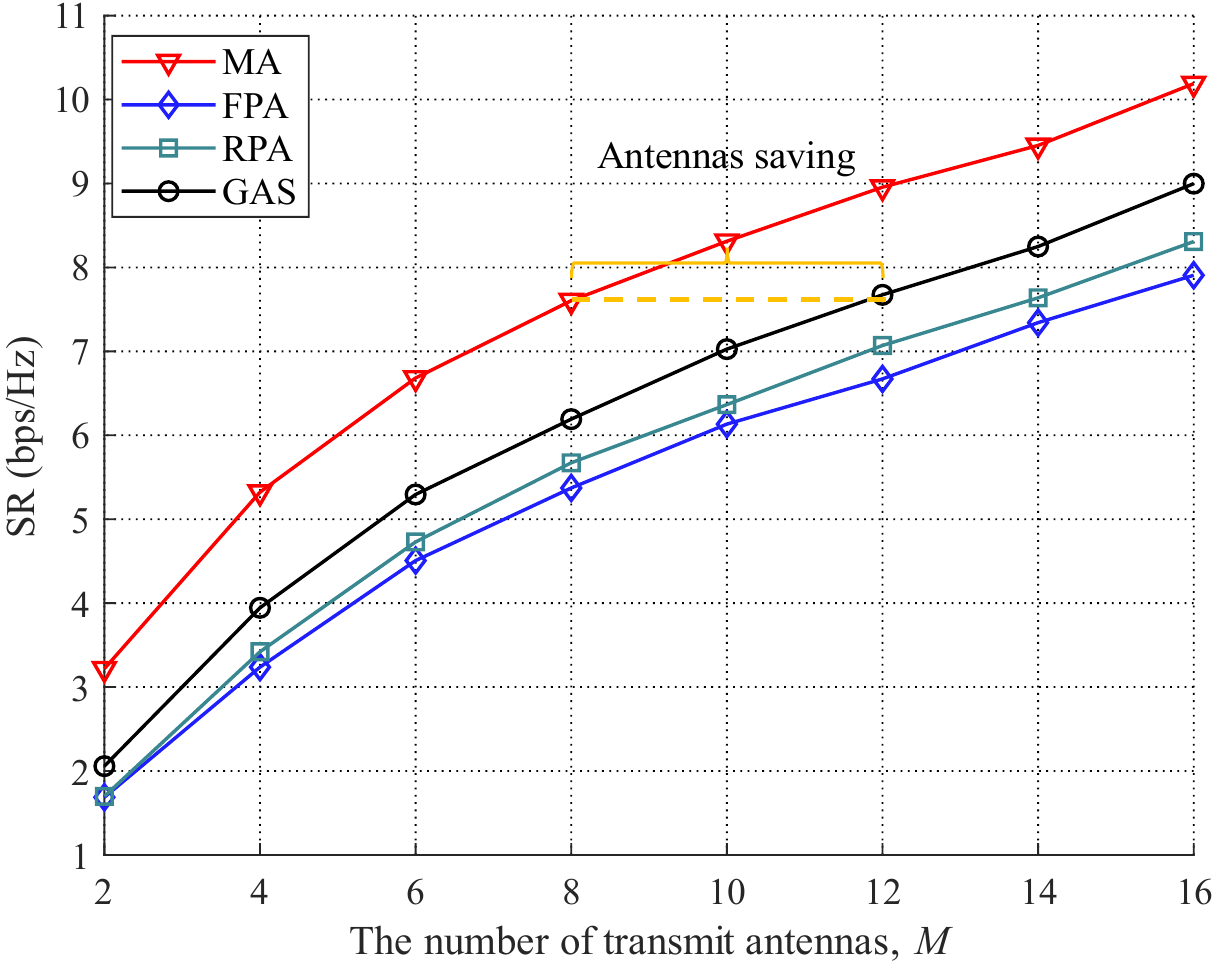}\vspace{-0.3cm}
	\caption{SR versus the number of transmit antennas.}
	\label{vs_M}
\end{figure}

\subsection{Baseline Schemes}
To fully demonstrate the significant advantages of MAs in improving the system's security performance, we propose the following four baseline schemes:
\begin{enumerate}
	\item {\bf{\text{FPA}}}: The BS is equipped with an FPA-based UPA with $M$ antennas spaced by $\frac{\lambda}{2}$.
	\item {\bf{\text{Random position antenna (RPA)}}}: The BS is equipped with $M$ antennas, where the antennas are random distributed in the transmit region $\mathcal{C}$, under the  constraint of the minimum distance $D$ between each other.  
	\item {\bf{\text{Exhaustive antenna selection (EAS)}}}: The BS is equipped with an FPA-based UPA with $2M$ antennas spaced by $\frac{\lambda}{2}$, where $M$ antennas are selected via the exhaustive search to maximize the SR. It should be pointed out that with the increase of $M$, the computational complexity rises sharply. For instance, when $M=16$, there are $C_{32}^{16}>6\times10^{8}$ selection schemes, which is hard to undertake. Thus, in the following simulation, we neglect this scheme when $M$ is large.
	\item {\bf{\text{Greedy antenna selection (GAS)}}}: The BS is equipped with an FPA-based UPA with $2M$ antennas spaced by $\frac{\lambda}{2}$. To avoid the heavy burden of calculations, we alternatively adopt the greedy antenna selection method. Specifically, we select the antenna that can maximize the current achieved SR step by step until the number of selected antennas reaches $M$. 
\end{enumerate}

\subsection{Performance Analysis Compared to Baseline Schemes}
{\itshape 1) Impact of the size of transmit region:} Fig. \ref{vs_A} shows the SR in different schemes versus the normalized size of transmit region. It is depicted that there is an obvious gain in SR of the MA-aided scheme as the size of transmit region increases, surpassing the performance of other baseline schemes. The reason is that a larger transmit region provides the MAs with more DoFs. Specifically, with a larger transmit region, the MAs can move to positions with improved channel condition. However, the performance gain brought by the larger transmit area is limited. When the transmit area exceeds $3\lambda \times 3\lambda$, the SR of the MA scheme keeps approximately constant with increasing $A$.

{\itshape 2) Impact of the number of transmit antennas:} Fig. \ref{vs_M} shows the SR versus the number of transmit antennas. We ignore the EAS scheme in this simulation owing to its high computational complexity. It is illustrated that the SR in all schemes increase a lot due to the spatial diversity gain and beamforming gain achieved by more antennas. Specifically, the SR in the MA scheme increases from 3.22 bps/Hz to 10.19 bps/Hz, achieving a performance gain of 216\%. Furthermore, it is demonstrated that the better spatial DoF utilization achieved by MAs can reduce the number of antennas required for the same level of security performance. Comparing the MA scheme with the GAS scheme, the former needs only 8 antennas while the latter requires $12\times2=24$ antennas to reach the SR threshold of 7.60 bps/Hz. Thus, the deployment of MAs is regarded as a promising technique to reduce the number of antennas in the next-generation communication systems.

\begin{figure}[t]
	\centering
	\includegraphics[width=3.25in]{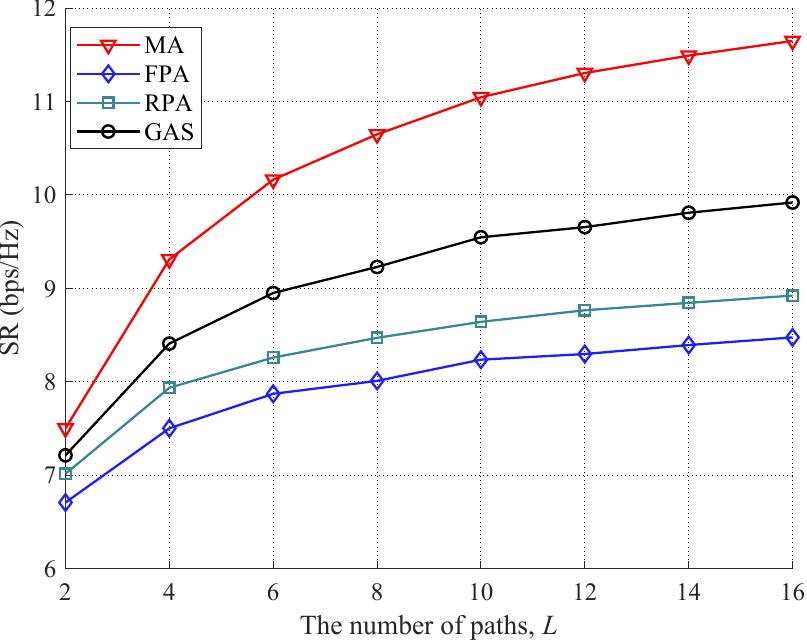}\vspace{-0.3cm}
	\caption{SR versus the number of paths.}
	\label{vs_L}
\end{figure}

{\itshape 3) Impact of the number of paths:} In order to fully exploit the influence of the number of paths on the MA scheme, we set $M=16$ and $A=8\lambda$. The EAS scheme is also not considered due to the high computational complexity. Fig. \ref{vs_L} illustrates the relationship between SR and the number of paths. We can observe that the SR increases in all schemes along with the increase in $L$ owing to the multi-path diversity. It is also illustrated that the MA scheme outperforms all the baseline schemes and the gaps grow larger with increasing $L$. Specifically, the MA scheme's performance gaps over the FPA, RPA and GAS schemes are 11.85\%, 6.95\% and 4.02\% when $L=2$, respectively, and those performance gaps increase to 37.46\%, 30.56\% and 17.44\% when $L=16$, respectively. It indicates that heavier small-scale fading generated by more paths may provide MAs with more DoFs and enhance the SR consequently.

\begin{figure}[t]
	\centering
	\includegraphics[width=3.25in]{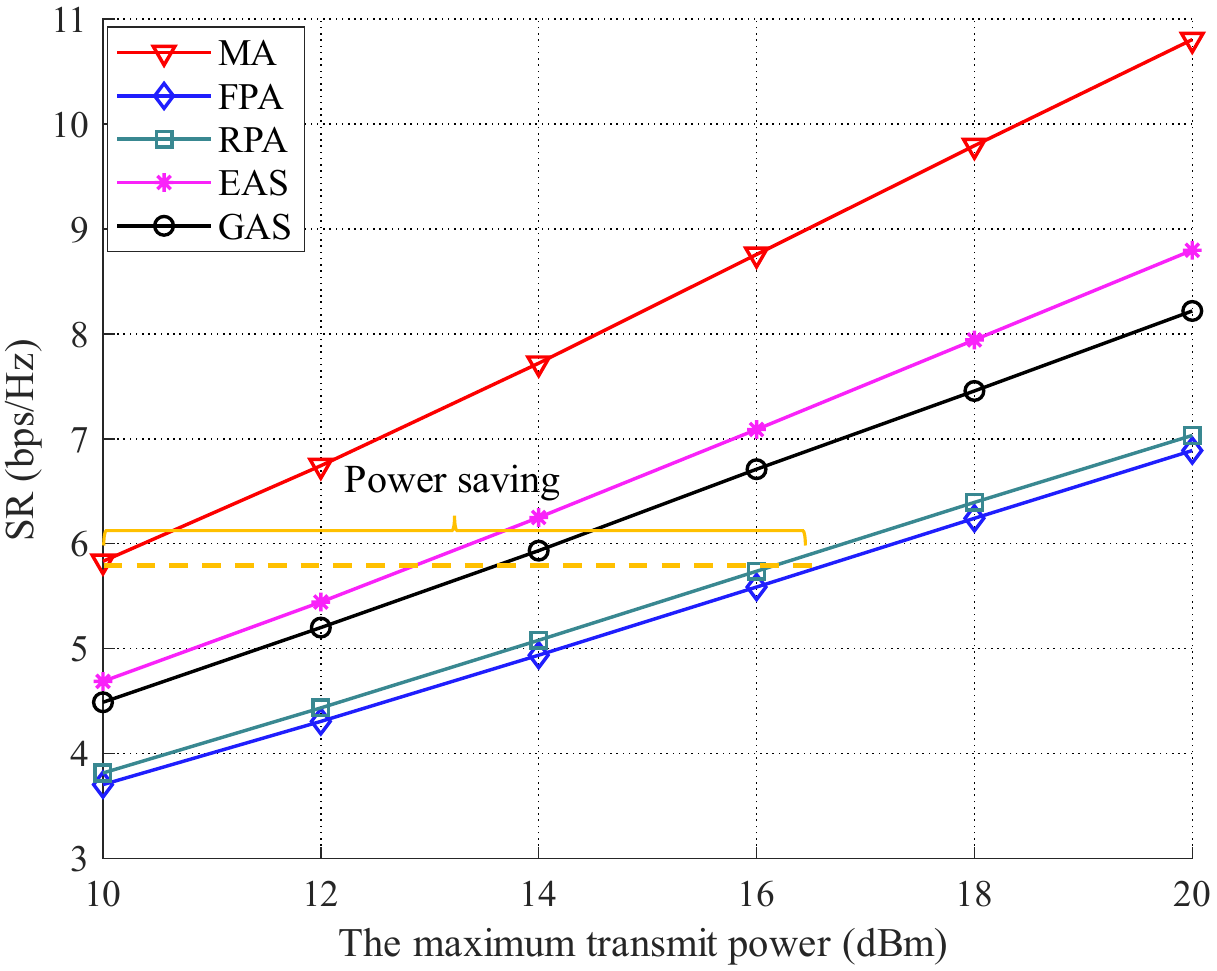}\vspace{-0.3cm}
	\caption{SR versus the maximum transmit power.}
	\label{vs_Pmax}
\end{figure}

{\itshape 4) Impact of the maximum transmit power:} Fig. \ref{vs_Pmax} shows the SR in different schemes versus the maximum transmit power. We can observe that with the increase in transmit power, the SR achieved by all schemes enhanced significantly and the MA scheme still overwhelms other baseline schemes. Specifically, when the transmit budget is 20 dBm, the MA scheme's performance gaps over the FPA, RPA, EAS and GAS are 56.87\%, 53.65\%, 22.82\% and 31.47\%, respectively. This is because the MAs can reconstruct the channel and improve the signal-to-interference-plus-noise-ratio (SINR) of the IR and consequently improve the security performance without a larger transmit power. Furthermore, comparing the MA scheme with the FPA scheme, the transmit power required by the MA scheme is much less than the latter to reach the same SR threshold. Specifically, when the achieved SR reaches 5.83 bps/Hz, only 10 dBm of transmit power is required by the MA scheme, whereas more than 16 dBm of transmit power is required by the FPA scheme.

{\itshape 5) Impact of the imperfect FRI:} In all the above discussions, we assume that the BS can estimate the channel perfectly. However, in practical application, the errors in channel estimation are unavoidable and will lead to performance degradation. Therefore, it is necessary to evaluate the MA-aided system's security performance under imperfect FRI\cite{xiao2023multiuser}. Specifically, we measure the FRI error from the following two aspects: AoD error and PRM error at the BS. For AoD error, let $\hat{\theta}_{k,j}^t$ and $\hat{\phi}_{k,j}^t$, $k\in\left\{I,E\right\}$ denote the estimated AoDs between the BS and the IR/Eve, and the differences between the estimated and the actual AoDs are assumed to be random variables obeying the uniform distribution, i.e., $\hat{\theta}_{k,j}^t-{\theta_{k,j}^t}\sim\mathcal{U}\left[-\frac{\mu}{2},\frac{\mu}{2}\right]$, $\hat{\phi}_{k,j}^t-{\phi_{k,j}^t}\sim\mathcal{U}\left[-\frac{\mu}{2},\frac{\mu}{2}\right]$, $1\le j\le L$, where $\mu$ denotes the maximum AoD error. For PRM error, let $\hat{\Sigma}_k={\rm{diag}}\left\{\hat{\sigma}_{1,1}^k,\dots,\hat{\sigma}_{L,L}^k\right\}$, $k\in\left\{I,E\right\}$ denote the estimated PRM from the BS to the IR/Eve, where the element $\hat{\sigma}_{l,l}^k$ is the estimated response coefficient between the $l$-th transmit path and the $l$-th receive path. Furthermore, the differences between the estimated and the actual response coefficients are assumed to be variables obeying the CSCG distribution, i.e., $\frac{\hat{\sigma}_{l,l}^k-\sigma_{l,l}^k}{\left|\sigma_{l,l}^k \right|}\sim\mathcal{CN}\left(0,\epsilon \right)$, $1\le l\le L$, where $\epsilon$ represents the variance of the normalized PRM error. 
\begin{figure}[t]
	\centering
	\includegraphics[width=3.25in]{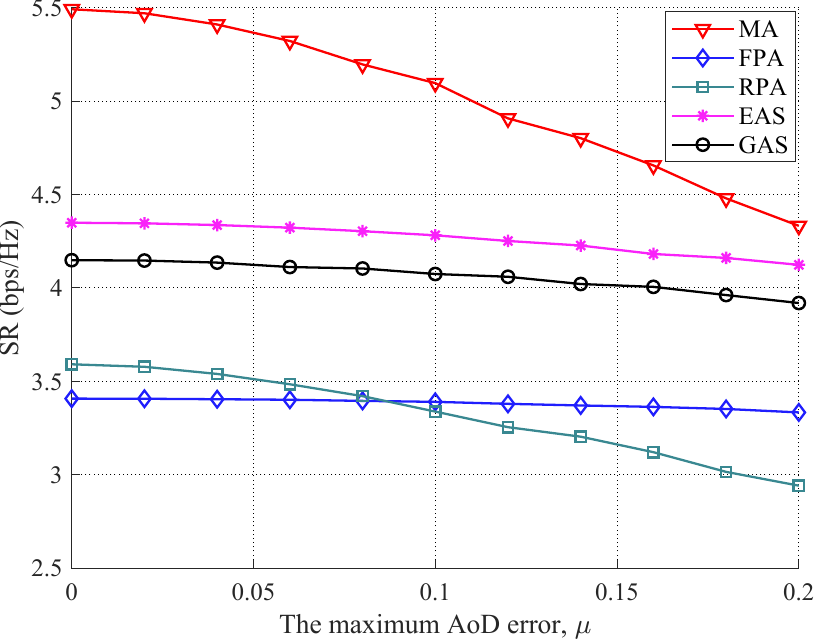}\vspace{-0.3cm}
	\caption{SR versus the maximum AoD error.}
	\label{vs_AoD_error}
\end{figure}
\begin{figure}[t]
	\centering
	\vspace*{4.5pt}
	\includegraphics[width=3.25in]{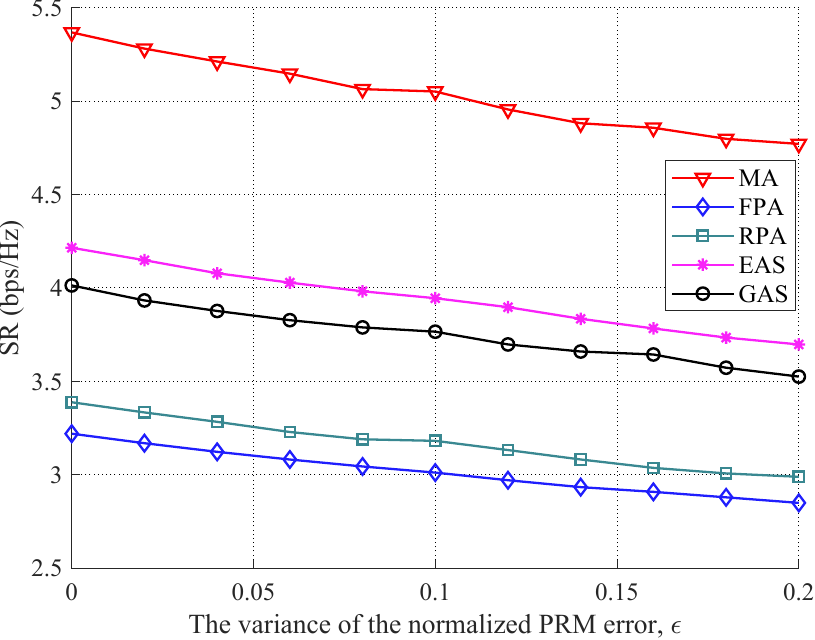}\vspace{-0.3cm}
	\caption{SR versus the variance of the normalized PRM error.}
	\label{vs_PRM_error}
\end{figure}

As shown in Fig. \ref{vs_AoD_error}, all the schemes achieve lower SR due to the AoD error, and the error leads to a severe performance degradation for the MA scheme compared to the other schemes. Specifically, the SR achieved in the MA scheme decreases from 5.49 bps/Hz to 4.33 bps/Hz, which  
suffers a performance degradation of 21.11\%. This is because the inaccurate AoD estimations may mislead the MAs to move to positions with extremely poor channel condition. Thus, accurate channel estimation is necessary for the MA-aided communication system. However, even with the maximum AoD error of 0.2 rad, the MA scheme still achieves better performance than other baseline schemes. 

Similarly, as shown in Fig. \ref{vs_PRM_error}, all the schemes achieve lower SR with the increase of PRM error. Specifically, the SR achieved in the MA scheme decreases from 5.37 bps/Hz to 4.77 bps/Hz,  suffering a performance degradation of 11.12\%. Likewise, the MA-aided scheme also prevails over the baseline schemes even when $\epsilon=0.2$.

\section{Conclusion}\label{Sec_conclusion}
In this paper, we proposed an MA-aided secure MIMO communication system, where the BS is equipped with MAs to enhance the system's security performance. We aimed at solving the SRM problem by jointly optimizing the TPC matrix, the AN covariance matrix and the positions of MAs. To address this challenge, a BCD-MM-based algorithm was proposed to tackle the non-convex SRM problem. Specifically, we first adopted the MMSE method to reformulate the problem and derived the optimal TPC matrix and the AN covariance matrix by applying the Lagrangian multiplier method. Then, we utilized the MM algorithm to obtain the optimal positions of the MAs. Simulation results demonstrated the significant advantages of the MA-aided system over conventional FPA system in enhancing the security performance of the system. Furthermore, it also illustrated the significance of accurate FRI for the security performance of the MA-aided systems.

\appendix
\subsection{Decouple ${\bf{g}}_{k}\left({\bf{t}}_m\right)$ from ${\bf{\xi}}\left({\bf{T}}\right)$}
Let us denote ${{\bf{a}}_{I,i}}\in \mathbb{C}^{L_I^t\times 1}$ as the $i$-th column of ${{\bf{A}}_I}$, $1 \le i \le M$, and ${{\bf{a}}_{E,i}}\in \mathbb{C}^{L_E^t\times 1}$ as the $i$-th column of ${{\bf{A}}_E}$, $1 \le i \le M$. Then, (\ref{other_items}) can be reformulated as 
\begin{subequations}
	\begin{align}
		&2{\rm {Re}}\left\lbrace {\rm{Tr}}\left({\bf{A}}_I{\bf{G}}_I^H \right)\right\rbrace \nonumber\\&=2{\rm {Re}}\left\lbrace   {\bf{g}}_I^H\left({\bf{t}}_m\right){\bf{a}}_{I, m} +\underbrace{\sum_{i=1,i\ne m}^M {\bf{g}}_I^H\left({\bf{t}}_i\right){\bf{a}}_{I, i} }_{\textrm {const }}\right\rbrace,\label{decoupled_first_term} \\ 
		&2{\rm {Re}}\left\lbrace {\rm{Tr}}\left({\bf{A}}_E{\bf{G}}_E^H \right)\right\rbrace \nonumber\\&=2{\rm {Re}}\left\lbrace   {\bf{g}}_E^H\left({\bf{t}}_m\right){\bf{a}}_{E, m} +\underbrace{\sum_{i=1,i\ne m}^M {\bf{g}}_E^H\left({\bf{t}}_i\right){\bf{a}}_{E, i} }_{\textrm {const }}\right\rbrace.\label{decoupled_third_term} 
	\end{align}
\end{subequations}
Similarly, we can decouple ${\bf{g}}_k\left({\bf{t}}_m\right)$ from (\ref{sum_of_3_6_fix}). We first let ${{\bf{V}}_X} = {{\bf{Q}}_X}{\bf{Q}}_X^H$, and denote the vector ${{\bf{q}}_{X,i}}\in \mathbb{C}^{M\times 1}$, where ${{\bf{q}}_{X,i}^H}$ is the $i$-th row of ${{\bf{Q}}_X}$, $1 \le i \le M$. In addition, we denote the vector ${{\bf{q}}_{E,i}}\in \mathbb{C}^{M\times 1}$, where ${{\bf{q}}_{E,i}^H}$ is the $i$-th row of ${{\bf{V}}_E}$, $1 \le i \le M$. Then, the terms of (\ref{sum_of_3_6_fix}) can be reformulated into the forms with decoupled ${\bf{g}}_k\left({\bf{t}}_m\right)$, which are shown at the top of Page 12. Following the above steps, we sum up all the terms as
\begin{figure*}[ht] 
	\centering
	\hrulefill 
	\vspace*{8pt} 
	\begin{equation}\label{decoupled_second_fourth_term_1}
		\begin{aligned}
			{\rm{Tr}}&\left( {\bf{G}}_I {\bf{V}}_X {\bf{G}}_I^H {\bf{C}}_I\right) 
			={\rm{Tr}}\left( {\bf{G}}_I {\bf{Q}}_X {\bf{Q}}_X^H {\bf{G}}_I^H {\bf{C}}_I\right) 
			={\rm{Tr}}\left[\left(\sum_{i=1}^M {\bf{g}}_I\left({\bf{t}}_i\right) {\bf{q}}_{X, i}^H\right)\left(\sum_{j=1}^M {\bf{q}}_{X, j} {\bf{g}}_I^H\left({\bf{t}}_j\right)\right) {\bf{C}}_I\right] \\
			=&\ {\rm{Tr}}\left[{\bf{g}}_I\left({\bf{t}}_m\right) {\bf{q}}_{X, m}^H {\bf{q}}_{X, m} {\bf{g}}_I^H\left({\bf{t}}_m\right) {\bf{C}}_I\right] 
			+{\rm{Tr}}\left[{\bf{g}}_I\left({\bf{t}}_m\right) {\bf{q}}_{X, m}^H\left(\sum_{j=1, j \neq m}^M {\bf{q}}_{X, j} {\bf{g}}_I^H\left({\bf{t}}_j\right)\right) {\bf{C}}_I\right] \\
			&+{\rm{Tr}}\left[\left(\sum_{i=1, i \neq m}^M {\bf{g}}_I\left({\bf{t}}_i\right) {\bf{q}}_{X, i}^H\right) {\bf{q}}_{X, m} {\bf{g}}_I^H\left({\bf{t}}_m\right) {\bf{C}}_I\right] 
			+\underbrace{{\rm{Tr}}\left[\left(\sum_{i=1, i \neq m}^M {\bf{g}}_I\left({\bf{t}}_i\right) {\bf{q}}_{X, i}^H\right)\left(\sum_{j=1, j \neq m}^M {\bf{q}}_{X, j} {\bf{g}}_I^H\left({\bf{t}}_j\right)\right) {\bf{C}}_I\right]}_{\textrm {const }},
		\end{aligned}
	\end{equation}
	\begin{equation}\label{decoupled_second_fourth_term_2}
		\begin{aligned}
			{\rm {Tr}}&\left( {\bf{G}}_E {\bf{V}}_X {\bf{G}}_E^H {\bf{C}}_X\right) 
			= {\rm {Tr}}\left( {\bf{G}}_E {\bf{Q}}_X {\bf{Q}}_X^H {\bf{G}}_E^H {\bf{C}}_X\right)  
			={\rm {Tr}}\left[\left(\sum_{i=1}^M {\bf{g}}_E\left({\bf{t}}_i\right) {\bf{q}}_{X, i}^H\right)\left(\sum_{j=1}^M {\bf{q}}_{X, j} {\bf{g}}_E^H\left({\bf{t}}_j\right)\right) {\bf{C}}_X\right] \\
			=&\ {\rm {Tr}}\left[{\bf{g}}_E\left({\bf{t}}_m\right) {\bf{q}}_{X, m}^H {\bf{q}}_{X, m} {\bf{g}}_E^H\left({\bf{t}}_m\right) {\bf{C}}_X\right] 
			+{\rm {Tr}}\left[{\bf{g}}_E\left({\bf{t}}_m\right) {\bf{q}}_{X, m}^H\left(\sum_{j=1, j \neq m}^M {\bf{q}}_{X, j} {\bf{g}}_E^H\left({\bf{t}}_j\right)\right) {\bf{C}}_X\right] \\
			& +{\rm {Tr}}\left[\left(\sum_{i=1, i \neq m}^M {\bf{g}}_E\left({\bf{t}}_i\right) {\bf{q}}_{X, i}^H\right) {\bf{q}}_{X, m} {\bf{g}}_E^H\left({\bf{t}}_m\right) {\bf{C}}_X\right] 
			+\underbrace{{\rm {Tr}}\left[\left(\sum_{i=1, i \neq m}^M {\bf{g}}_E\left({\bf{t}}_i\right) {\bf{q}}_{X, i}^H\right)\left(\sum_{j=1, j \neq m}^M {\bf{q}}_{X, j} {\bf{g}}_E^H\left({\bf{t}}_j\right)\right) {\bf{C}}_X\right]}_{\textrm{const}},
		\end{aligned}
	\end{equation}
	\begin{equation}\label{decoupled_second_fourth_term_3}
		\begin{aligned}
			{\rm {Tr}}&\left( {\bf{G}}_E {\bf{V}}_E {\bf{V}}_E^H {\bf{G}}_E^H {\bf{C}}_E\right) 
			={\rm {Tr}}\left[\left(\sum_{i=1}^M {\bf{g}}_E\left({\bf{t}}_i\right) {\bf{q}}_{E, i}^H\right)\left(\sum_{j=1}^M {\bf{q}}_{E, j} {\bf{g}}_E^H\left({\bf{t}}_j\right)\right) {\bf{C}}_E\right] \\
			=&\ {\rm{Tr}}\left[{\bf{g}}_E\left({\bf{t}}_m\right) {\bf{q}}_{E, m}^H {\bf{q}}_{E, m} {\bf{g}}_E^H\left({\bf{t}}_m\right) {\bf{C}}_E\right] 
			+{\rm {Tr}}\left[{\bf{g}}_E\left({\bf{t}}_m\right) {\bf{q}}_{E, m}^H\left(\sum_{j=1, j \neq m}^M {\bf{q}}_{E, j} {\bf{g}}_E^H\left({\bf{t}}_j\right)\right) {\bf{C}}_E\right] \\
			&+{\rm {Tr}}\left[\left(\sum_{i=1, i \neq m}^M {\bf{g}}_E\left({\bf{t}}_i\right) {\bf{q}}_{E, i}^H\right) {\bf{q}}_{E, m} {\bf{g}}_E^H\left({\bf{t}}_m\right) {\bf{C}}_E\right] 
			+\underbrace{{\rm {Tr}}\left[\left(\sum_{i=1, i \neq m}^M {\bf{g}}_E\left({\bf{t}}_i\right) {\bf{q}}_{E, i}^H\right)\left(\sum_{j=1, j \neq m}^M {\bf{q}}_{E, j} {\bf{g}}_E^H\left({\bf{t}}_j\right)\right) {\bf{C}}_E\right]}_{\textrm{const}}.
		\end{aligned}
	\end{equation}
	\hrulefill
\end{figure*}

\begin{equation}
	{\xi}\left({\bf{T}}\right)=-\textrm{(\ref{decoupled_first_term})}-\textrm{(\ref{decoupled_third_term})}+\textrm{(\ref{decoupled_second_fourth_term_1})}+\textrm{(\ref{decoupled_second_fourth_term_2})}+\textrm{(\ref{decoupled_second_fourth_term_3})},
\end{equation}
and consequently decouple the $m$-th MA's FRV ${\bf{g}}_k\left({\bf{t}}_m\right)$ from ${\xi}\left({\bf{T}}\right)$.

\subsection{Calculate $\nabla \gamma \left( {{{\bf{t}}_m}} \right)$ and ${\nabla ^2}\gamma \left( {{{\bf{t}}_m}} \right)$}
Let us denote the $j$-th element of ${\boldsymbol{\eta }}_I^i$ as ${\eta_{I,j}}$, and the $j$-th element of ${\boldsymbol{\eta }}_E^i$ as ${\eta_{E,j}}$, then $\gamma\left({\bf{t}}_m\right)$ can be rewritten as\cite{10243545}
\begin{equation}
	\begin{aligned}
		\gamma (& {{{\bf{t}}_m}} ) \\=&\  2\left(\sum\limits_{j = 1}^{L_I^t} {\left| \eta_{I,j} \right|\cos \left({\kappa _{I,j}}\left({{\bf{t}}_m}\right)\right) + } \sum\limits_{j = 1}^{L_E^t} {\left| \eta_{E,j} \right|\cos \left({\kappa _{E,j}}\left({{\bf{t}}_m}\right)\right)}\right)\\
		&+ \textrm{const},
	\end{aligned}
\end{equation}
where
\begin{equation}
	{\kappa _{I,j}}\left({{\bf{t}}_m}\right) = \frac{{2\pi }}{\lambda }\rho _{I,j}^t\left({{\bf{t}}_m}\right) - \angle \eta_{I,j},
\end{equation}
\begin{equation}
	\rho _{I,j}^t\left({{\bf{t}}_m}\right) = x_m^t\sin \theta _{I,j}^t\cos \phi _{I,j}^t + y_m^t\cos \theta _{I,j}^t,
\end{equation}
\begin{equation}
	{\kappa _{E,j}}\left({{\bf{t}}_m}\right) = \frac{{2\pi }}{\lambda }\rho _{E,j}^t\left({{\bf{t}}_m}\right) - \angle \eta_{E,j},
\end{equation}
\begin{equation}
	\rho _{E,j}^t\left({{\bf{t}}_m}\right) = x_m^t\sin \theta _{E,j}^t\cos \phi _{E,j}^t + y_m^t\cos \theta _{E,j}^t,
\end{equation}
respectively. The gradient vector and the Hessian matrix of $\gamma\left({\bf{t}}_m\right)$ w.r.t. ${\bf{t}}_m$ can be represented as  $\nabla \gamma \left( {{{\bf{t}}_m}} \right) = {\left[ {\frac{{\partial \gamma \left( {{{\bf{t}}_m}} \right)}}{{\partial x_m^t}},\frac{{\partial \gamma \left( {{{\bf{t}}_m}} \right)}}{{\partial y_m^t}}} \right]^T}$ and $\nabla^2 \gamma\left({\bf{t}}_m\right)=\left[\begin{array}{cc}
	\frac{\partial^2 \gamma\left({\bf{t}}_m\right)}{\partial x_m^t \partial x_m^t} & \frac{\partial^2 \gamma\left({\bf{t}}_m\right)}{\partial x_m^t \partial y_m^t} \\
	\frac{\partial^2 \gamma\left({\bf{t}}_m\right)}{\partial y_m^t \partial x_m^t} & \frac{\partial^2 \gamma\left({\bf{t}}_m\right)}{\partial y_m^t \partial y_m^t}
\end{array}\right] $, and the relative terms are shown at the top of Page 13.
\begin{figure*}[ht] 
	\centering
	\hrulefill 
	\vspace*{8pt} 
	\begin{subequations}\label{first_order_grad}
		\begin{align}
			& \frac{\partial \gamma\left({\bf{t}}_m\right)}{\partial x_m^t}=-\frac{4 \pi}{\lambda}\left(\sum_{j=1}^{I_I^t}\left|\eta_{I, j}\right| \sin \theta_{I, j}^t \cos \phi_{I, j}^t \sin \left(\kappa_{I, j}\left({\bf{t}}_m\right)\right)+\sum_{j=1}^{L_E^t}\left|\eta_{E, j}\right| \sin \theta_{E, j}^t \cos \phi_{E, j}^t \sin \left(\kappa_{E, j}\left({\bf{t}}_m\right)\right)\right), \\
			& \frac{\partial \gamma\left({\bf{t}}_m\right)}{\partial y_m^t}=-\frac{4 \pi}{\lambda}\left(\sum_{j=1}^{L_I^t}\left|\eta_{I, j}\right| \cos \theta_{I, j}^t \sin \left(\kappa_{I, j}\left({\bf{t}}_m\right)\right)+\sum_{j=1}^{L_E^t}\left|\eta_{E, j}\right| \cos \theta_{E, j}^t \sin \left(\kappa_{E, j}\left({\bf{t}}_m\right)\right)\right),
		\end{align}
	\end{subequations}
	\begin{subequations}\label{second_order_hessen}
		\begin{align}
			& \frac{\partial^2 \gamma\left({\bf{t}}_m\right)}{\partial x_m^t \partial x_m^t}=-\frac{8 \pi^2}{\lambda^2}\left(\sum_{j=1}^{L_t^t}\left|\eta_{I, j}\right| \sin ^2 \theta_{I, j}^t \cos ^2 \phi_{I, j}^t \cos \left(\kappa_{I, j}\left({\bf{t}}_m\right)\right)+\sum_{j=1}^{L_E^t}\left|\eta_{E, j}\right| \sin ^2 \theta_{E, j}^t \cos ^2 \phi_{E, j}^t \cos \left(\kappa_{E, j}\left({\bf{t}}_m\right)\right)\right), \\
			& \frac{\partial^2 \gamma\left({\bf{t}}_m\right)}{\partial x_m^t \partial y_m^t}=-\frac{8 \pi^2}{\lambda^2}\left(\sum_{j=1}^{L_I^t}\left|\eta_{I, j}\right| \sin \theta_{I, j}^t \cos \phi_{I, j}^t \cos \theta_{I, j}^t \cos \left(\kappa_{I, j}\left({\bf{t}}_m\right)\right)+\sum_{j=1}^{L_E^t}\left|\eta_{E, j}\right| \sin \theta_{E, j}^t \cos \phi_{E, j}^t \cos \theta_{E, j}^t \cos \left(\kappa_{E, j}\left({\bf{t}}_m\right)\right)\right), \\
			& \frac{\partial^2 \gamma\left({\bf{t}}_m\right)}{\partial y_m^t \partial x_m^t}=-\frac{8 \pi^2}{\lambda^2}\left(\sum_{j=1}^{L_I^t}\left|\eta_{I, j}\right| \sin \theta_{I, j}^t \cos \phi_{I, j}^t \cos \theta_{I, j}^t \cos \left(\kappa_{I, j}\left({\bf{t}}_m\right)\right)+\sum_{j=1}^{L_E^t}\left|\eta_{E, j}\right| \sin \theta_{E, j}^t \cos \phi_{E, j}^t \cos \theta_{E, j}^t \cos \left(\kappa_{E, j}\left({\bf{t}}_m\right)\right)\right), \\
			& \frac{\partial^2 \gamma\left({\bf{t}}_m\right)}{\partial y_m^t \partial y_m^t}=-\frac{8 \pi^2}{\lambda^2}\left(\sum_{j=1}^{L_I^t}\left|\eta_{I, j}\right| \cos ^2 \theta_{I, j}^t \cos \left(\kappa_{I, j}\left({\bf{t}}_m\right)\right)+\sum_{j=1}^{L_E^t}\left|\eta_{E, j}\right| \cos ^2 \theta_{E, j}^t \cos \left(\kappa_{E, j}\left({\bf{t}}_m\right)\right)\right).
		\end{align}
	\end{subequations}
	\hrulefill
\end{figure*}

\subsection{Construct ${\delta _m}$}
Since 
\begin{equation}
	\begin{aligned}
		\left\| {{\nabla ^2}\gamma \left( {{{\bf{t}}_m}} \right)} \right\|_2^2 \le& \left\| {{\nabla ^2}\gamma \left( {{{\bf{t}}_m}} \right)} \right\|_F^2 \\
		\le& 4{\left( {\frac{{8{\pi ^2}}}{{{\lambda ^2}}}\left( {\sum\limits_{j = 1}^{L_I^t} {\left| {{\eta_{I,j}}} \right|}  + \sum\limits_{j = 1}^{L_E^t} {\left| {{\eta_{E,j}}} \right|} } \right)} \right)^2},
	\end{aligned}
\end{equation}
and
\begin{equation}
	{\left\| {{\nabla ^2}\gamma \left( {{{\bf{t}}_m}} \right)} \right\|_2}{{\bf{I}}_2}\succeq {\nabla ^2}\gamma \left( {{{\bf{t}}_m}} \right),
\end{equation}
we can construct $\delta_m$ as
\begin{equation}\label{value of delta_m}
	{\delta _m} = \frac{{16{\pi ^2}}}{{{\lambda ^2}}}\left( {\sum\limits_{j = 1}^{L_I^t} {\left| {{\eta_{I,j}}} \right|}  + \sum\limits_{j = 1}^{L_E^t} {\left| {{\eta_{E,j}}} \right|} } \right).
\end{equation}
It can be derived that
\begin{equation}
	{\delta _m}{{\bf{I}}_2}\succeq{\left\| {{\nabla ^2}\gamma \left( {{{\bf{t}}_m}} \right)} \right\|_2}{{\bf{I}}_2}\succeq{\nabla ^2}\gamma \left( {{{\bf{t}}_m}} \right),
\end{equation}
thus the construction of $\delta_m$ is completed as in \cite{10243545}.

\bibliographystyle{IEEEtran}
\bibliography{Reference}

\end{document}